\documentclass[12pt]{article}
\usepackage{amsmath,amssymb}
\usepackage{amsthm}
\usepackage{mathrsfs}
\usepackage{cite}
\usepackage[top=3cm,bottom=3cm, left=2.3cm,right=2.3cm]{geometry}
\numberwithin{equation}{section}

\newcommand{\co}{{\rm c}}

\newcommand{\na}{\nabla}
\newcommand{\hph}{\hphantom}
\renewcommand{\b}{\bar}
\renewcommand{\d}{\dot}

\newcommand{\df}{\dfrac}
\newcommand{\ul}{\underline}
\newcommand{\der}{\partial}

\renewcommand{\(}{\left(}
\renewcommand{\)}{\right)}

\newcommand{\dg}{\dagger}
\newcommand{\wed}{\wedge}
\newcommand{\bmx}{\left(\begin{matrix}}
\newcommand{\emx}{\end{matrix}\right)}

\newcommand{\corr}{\leftrightarrow}

\newcommand{\stog}[4]{
\bmx
 0 & (\sigma_{#1}^{#2})_{{#3}\dot{#4}} \\
(\bar{\sigma}_{#1}^{#2})^{\dot{#3}{#4}} & 0
\emx 
}

\newdimen\Tdim
\newdimen\Ddim
\def\Tspan#1{{\setbox0=\hbox{$#1$}%
\Tdim\ht0\advance\Tdim\dp0\advance\Tdim.7ex\Ddim\dp0\advance\Ddim.4ex\rule[-\Ddim]{0pt}{\Tdim}\box0}}

\begin{document}
\begin{titlepage}
\hfill WU-HEP-16-11
\begin{center}
 {\Large\bf
Abelian tensor hierarchy in 
\\[3mm]
4D ${\cal N}=1$ 
conformal supergravity
}
\vspace{12mm}

Shuntaro Aoki${}^1$\footnote{
email: shun-soccer@akane.waseda.jp},
Tetsutaro Higaki${}^{2,3}$\footnote{
email: thigaki@rk.phys.keio.ac.jp},
Yusuke Yamada${}^{2,3}$\footnote{
email: yusuke-yamada@keio.jp}
 and 
Ryo Yokokura${}^2$\footnote{
email: ryokokur@rk.phys.keio.ac.jp}
\\
\vspace{5mm}

${}^1${\small\it Department of Physics,
Waseda  University,
 Tokyo 169-8555, Japan
}
\\
${}^2${\small\it Department of Physics,
Keio  University,
Yokohama 223-8522, Japan
}
\\
${}^3${\small\it 
Research and Education Center for Natural Sciences,
Keio  University,
Yokohama 223-8521, Japan
}

\vspace{15mm}

\end{center}

\begin{abstract}
\noindent
We consider Abelian tensor hierarchy in four-dimensional 
${\cal N}=1$ 
supergravity in the conformal superspace formalism, where
the so-called covariant approach is used to antisymmetric tensor fields.
We introduce $p$-form gauge superfields 
as superforms in the conformal superspace.
We solve the Bianchi identities under the constraints
for the superforms.
As a result, 
each of form fields is
expressed by 
a single 
gauge invariant superfield.
The action of superforms is shown with the invariant 
superfields.
We also show the relation between the superspace formalism
and
the superconformal tensor calculus.
\end{abstract}
\end{titlepage}
\setcounter{footnote}{0}

\section{Introduction}

The superstring theory is regarded as a candidate for the unified theory of 
fundamental interactions including gravity. The theory contains 
strings/branes, which can describe gravity, gauge fields and matter ones in the low energy limit.
For stable branes to describe our universe, anti-symmetric tensor fields are coupled to their preserved charges because they are extended objects. Further, unstable tachyons are avoided in the presence of the supersymmetry (SUSY).
As a consequence, one may take
supergravity (SUGRA), in which
there exist such tensor fields on top of gravity, as a plausible low energy effective theory 
of the superstring theory.

Four-dimensional (4D) 
effective action is obtained through a compactification of extra dimensions.
Hence, 4D ${\cal N}=1$ SUGRA with tensor fields is a possible starting
point 
to construct the effective description of the superstring theory,
because ${\cal N}=1$ SUGRA is a chiral theory, 
which has to contain the Standard Model
of particle physics.

For these reasons,
our interest is to construct 
the effective models of the superstring
including $p$-form gauge fields within 4D
 ${\cal N}=1$ SUGRA~\cite{{Siegel:1979ai},{Gates:1980ay},{Cecotti:1987qr},
{Binetruy:1996xw},{Ovrut:1997ur}}.
For the effective description of the superstring theory,
we need to respect the structure of 
ten-dimensional antisymmetric tensors.
In 4D effective theory, such antisymmetric tensors and their gauge
transformations are described by the 4D form fields, whose
transformations inevitably contain form fields with different
ranks.
Such a structure is called a tensor hierarchy~\cite{{deWit:2005hv},{Hartong:2009az},
{bib:BBLR},{Becker:2016rku}}, 
and is related to the anomaly cancellation conditions in the string theory.
Therefore, the construction of 
a tensor hierarchy in 4D ${\cal N}=1$ SUGRA is 
desirable in the context of string models.

In this paper, we consider the tensor hierarchy in 4D ${\cal N}=1$ SUGRA.
In particular, we focus on the construction
 which 
is inspired by structures of 
the geometries of extra dimensions.
Becker et al. did such a construction in 4D 
${\cal N}=1$ global SUSY~\cite{bib:BBLR}.
Our 4D SUGRA description can be applicable to discuss the
 roles of antisymmetric tensors, e.g. in cosmology~\cite{{Kaloper:2008fb},{Kaloper:2011jz},{Ferrara:2013rsa},{Kaloper:2014zba},{Farakos:2014gba},{Dudas:2014pva}} and SUSY breaking~\cite{Burgess:1995kp,Binetruy:1995hq,Farakos:2016hly}.

We will use the conformal superspace formalism~\cite{bib:B1},
which
is a superspace formalism of conformal
supergravity. 
It has larger gauge symmetries than the superconformal tensor 
calculus~\cite{{bib:KTVN},{bib:KT},{bib:TVN},{bib:FGVN},{bib:CFGVP},{bib:KUigc},{Kugo:1982cu},{bib:KU},{bib:KKLVP}}
and  Poincar\'e superspace~\cite{{bib:WB},{bib:BG2}}.
The symmetries will be useful to 
construct the SUGRA system coupled to the tensors and matters.
We can straightforwardly reproduce the corresponding system 
in terms of the Poincar\'e superspace 
and also the superconformal tensor calculus 
due to their correspondences~\cite{{bib:B1},{bib:KYY}}.

We will adopt the 
so-called covariant approach~\cite{{Binetruy:1996xw},{bib:WB},{bib:BG2},{bib:SS},{Muller:1985vga}}.
In this approach, we regard bosonic tensors 
as  components of  differential superforms in superspace.
This makes the local SUSY properties of the tensors manifest.
In particular, it is straightforward to obtain 
gauge invariant superfields including 
bosonic field strengths.

This paper is organized as follows.
In Sec.~\ref{sec:ath}, we review
4D ${\cal N}=1$ conformal superspace briefly.
Then, Abelian tensor hierarchy is introduced to 
the conformal superspace.
Section~\ref{sec:cBi} is devoted to 
 impose constraints on field strength, 
and to show the solutions to the Bianchi identities.
In Sec.~\ref{sec:cf}, we present the component formalism,
which is written by superconformal 
tensor calculus.
In Sec.~\ref{sec:sca}, 
a superconformally invariant action 
for $p$-form gauge superfields
is proposed.
We conclude this paper in Sec.~\ref{sec:concl}.
In appendix~\ref{sec:not},  our notations are
summarized. 
We present the explicit derivations 
 of the solutions of Bianchi identities in
 appendix~\ref{sec:der}.
We show the explicit forms of bosonic field strengths 
in appendix~\ref{sec:bfs}.
Throughout this paper, the terms 
``form'', ``gauge field'' and ``field strength''
 are used to refer
``superform'', ``gauge superfield'' and ``field strength superfield'',
respectively.
We use the conventions of Ref.~\cite{bib:KYY} for 
conformal superspace except the notation of torsion.
We use $T_{CB}{}^A$ to refer the torsion, which is 
equal to $R(P)_{CB}{}^A$ in Ref.~\cite{bib:KYY}.
We also use the convention of Ref.~\cite{bib:BG2} for 
superforms, exterior derivatives and interior products.
\section{Abelian tensor hierarchy in conformal superspace}
\label{sec:ath}
In this section, we introduce the Abelian tensor hierarchy 
into conformal superspace.
We begin with a brief review of conformal superspace.
Abelian tensor hierarchy is then introduced into conformal
superspace.

\subsection{Conformal superspace}
Conformal SUGRA is one of the most convenient 
formulation of SUGRA thanks to its larger gauge symmetries.
Conformal superspace is a superspace approach to formulate conformal 
SUGRA.
In the conformal superspace, we can formulate conformal SUGRA 
in a geometrical manner.

Conformal SUGRA is constructed as the gauge theory
 of superconformal group.
We formulate conformal SUGRA in a superspace~\cite{bib:B1}.
Superspace is a space where the coordinates are spanned by 
ordinary bosonic spacetime coordinates $x^m$ and 
fermionic coordinates $(\theta^\mu,\b\theta_{\d\mu})$. 
Here, $m,n,...$ are used for curved vector indices, 
$\mu,\nu,...$ 
for undotted spinor indices, and $\d\mu,\d\nu,...$ for
dotted spinor indices.
In the superspace, we can deal with bosonic translations and 
SUSY transformations at the same time,
 since SUSY transformations are understood as 
fermionic translations.
Thus, we denote both bosonic and fermionic coordinates 
as $z^M=(x^m,\theta^\mu,\b\theta_{\d\mu})$, where capital Roman letters
 $M,N,...$ express the sets of curved vector and spinor indices.

Conformal superspace is a superspace 
where gauge fields of the superconformal symmetry
are introduced.
The generators 
of the superconformal group are 
the following elements: spacetime translations $P_a$, 
SUSY transformations $Q_{\ul{\alpha}}$, 
Lorentz transformations $M_{ab}$, 
dilatation $D$, chiral rotation $A$,
conformal boosts $K_a$ 
and conformal SUSY transformations $S_{\ul{\alpha}}$.
Here, Roman letters $a,b,...$ denote flat vector indices,
Greek letters $\alpha,\beta,...$ and $\d\alpha,\d\beta,...$ 
express
undotted and dotted flat spinor indices, respectively.
$\ul{\alpha},\ul{\beta},...$ denote both spinor indices 
$\ul{\alpha}=(\alpha,\d\alpha)$.
In the superspace, we can deal with bosonic translations and 
SUSY transformations at the same time.
Therefore, we simply write $P_A$ and $K_A$ as 
$P_A=(P_a, Q_\alpha , \b{Q}^{\d\alpha})$
and 
$K_A=(K_a,S_\alpha,\b{S}^{\d\alpha})$, respectively.
Here, we use Roman capital indices $A,B,...$ for the sets of 
Lorentz vectors and spinors $A=(a,\alpha,\d\alpha)$.
All the generators of superconformal group are 
denoted $X_{\cal A}=(P_A, M_{ab}, D, A, K_A)$,
where the calligraphic letters ${\cal A},{\cal B},...$ are 
used to refer the indices of the generators of the superconformal 
group.

The gauge fields of the superconformal symmetry are defined as
\begin{equation}
 h_M{}^{\cal A} X_{\cal A}
=
E_M{}^A P_A+\df{1}{2}\phi_M{}^{ab}M_{ba}
+B_M D+A_M A+f_M{}^A K_A,
\end{equation}
where $E_M{}^A$ is the vielbein, 
$\phi_M{}^{ab}$ is the spin connection,
$B_M$, $A_M$ and $f_M{}^A$ are the gauge 
fields corresponding to $D$, $A$ and $K_A$,
respectively.
We assume that $E_M{}^A$ are invertible,
and their inverses are denoted as $E_A{}^M$:
\begin{equation}
 E_M{}^A E_A{}^N=\delta_M{}^N,
\quad
E_A{}^N E_N{}^B=\delta_A{}^B.
\end{equation}
Using a differential form \cite{bib:BG2}, 
the gauge fields are expressed as
\begin{equation}
 h^{\cal A} =dz^M h_M{}^{\cal A}.
\end{equation}

The gauged superconformal transformations
are generated by $\delta_G(\xi^{\cal A}X_{\cal A})$.
Here, $\xi^{\cal{A}}$ are real parameter superfields,
and $\xi^{\cal{A}}X_{\cal{A}}$ denotes
\begin{equation}
\xi^{\cal{A}}X_{\cal{A}}
=\xi(P)^AP_A+\df{1}{2}\xi(M)^{ab}M_{ba}+\xi(D) D+\xi(A)A+{\xi}(K)^{A}K_A.
\end{equation}
 The gauge fields $h_M{}^{\cal A}$ receive the superconformal transformations
$\delta_G(\xi^{{\cal A}'}X_{{\cal A}'})$ as
\begin{equation}
\delta_G(\xi^{{\cal B}'}X_{{\cal B}'}) {h_M}^{{\cal A}}
=\der_M \xi^{{\cal B}'}{\delta_{{\cal B}'}}^{{\cal A}}
+{h_M}^{{\cal C}}\xi^{{\cal B}'}{f_{{\cal B'C}}}^{\cal A},
\end{equation}
where primed calligraphic indices ${\cal A}',{\cal B'},...$
mean all the
superconformal generators except for $P_A$:
$X_{\cal A'}=(M_{ab},D,A,K_A)$.

We shall define SUSY transformations and spacetime translations. 
In the superspace approach, we can deal with SUSY transformations and 
spacetime translations at the same time, namely $P_A$-transformations.
We relate $P_A$-transformations 
to the general coordinate transformation
$\delta_{\text{GC}}$ 
by using field-independent parameter superfields $\xi^A$ as
\begin{equation}
\delta_G(\xi^AP_A) = 
\delta_{\text{GC}}(\xi^M)
-\delta_G(\xi^M{h_M}^{{\cal B}'}X_{{\cal B}'}),
\end{equation}
where $\xi(P)^A$ are abbreviated to $\xi^A$, and 
$\xi^M $ are defined by $\xi^M:=\xi^AE_A{}^M$.
The $P_A$-transformations acting on a superfield $\Phi$ with
no curved index define the 
superconformally covariant derivatives $\na_A$ as
\begin{equation}
\delta_G(\xi^AP_A)\Phi=\xi^AP_A\Phi
=\xi^A\na_A\Phi 
 = \xi^M\na_M\Phi
=\xi^M(\der_M-{h_M}^{{\cal A}'}X_{{\cal A}'})\Phi.
\end{equation}
Let us consider the curvatures associated with 
the superconformal symmetry, which appear in the Bianchi identities.
They are defined by
\begin{equation}
{R_{MN}}^{{\cal A}} = \der_M{h_N}^{{\cal A}}-\der_N{h_M}^{{\cal A}}
-({E_N}^{C}{h_M}^{{\cal B}'}-{E_M}^{C}{h_N}^{{\cal B}'})
{f_{{\cal B}'C}}^{{\cal A}} -
{h_N}^{{\cal C}'}{h_M}^{{\cal B}'}{f_{{\cal B'C'}}}^{{\cal A}}.
\end{equation}
Here, we use the convention of ``implicit grading''
\cite{bib:B1}.
Using differential forms, they are expressed as
\begin{equation}
R^{\cal A}
=\df{1}{2}dz^M\wed dz^N {R_{NM}}^{\cal A} 
= d h^{\cal A}
-E^{B}\wed h^{\cal C'}
{f_{{\cal C'}B}}^{\cal A} -
\df{1}{2}
h^{\cal B'}\wed h^{\cal C'}{f_{\cal C'B'}}^{\cal A},
\end{equation}
where $E^A=dz^M E_M{}^A$.
In particular, the curvatures associated with $P_A$ are 
the torsion two-forms $T^A$.
The torsions are given by explicitly
\begin{equation}
 T^A=dE^A-E^C\wed h^{{\cal B}'}f_{{\cal B}'C}{}^A
=\df{1}{2}E^B\wed E^C T_{CB}{}^A,
\label{eq:tor}
\end{equation}
which appear in the Bianchi identities for 
$p$-form gauge fields discussed later.
The curvatures are expressed also in terms of the 
(anti-)commutation relations of the 
superconformally
 covariant derivatives 
\begin{equation}
 [\na_A,\na_B]
=-R_{AB}{}^{\cal C}X_{\cal C}.
\end{equation}
The curvatures are constrained so that 
(anti-)commutation relations of the covariant derivatives
are given by
\begin{equation}
 \{\na_\alpha,\na_\beta\}=0,\quad
\{\b\na_{\d\alpha},\b\na_{\d\beta}\}=0,\quad
\{\na_\alpha,\b\na_{\d\beta}\}=-2i \na_{\alpha\d\beta},
\end{equation}
\begin{equation}
\begin{split}
&[\na_\alpha,\na_{\beta\d\gamma}]
=-2i\epsilon_{\alpha\beta}{\cal W}_{\d\gamma},  \qquad
[\b\na_{\d\alpha},\na_{\d\beta\gamma}]
=-2i\epsilon_{\d\alpha\d\beta}{\cal W}_\gamma, 
\\
&[\na_{\alpha\d\alpha},\na_{\beta\d\beta}] 
=
 \epsilon_{\d\alpha\d\beta}
\{\na_{(\alpha},{\cal W}_{\beta)}\}+\epsilon_{\alpha\beta}
\{\b\na_{(\d\alpha},{\cal W}_{\d\beta)}\}, 
\end{split}
\end{equation}
where 
\begin{equation}
 \begin{split}
{\cal W}_\alpha &=
(\epsilon\sigma^{bc})^{\beta\gamma}W_{\alpha\beta\gamma}M_{cb}
+\df{1}{2}\(\na^{\gamma}{W_{\gamma\alpha}}^\beta\)S_\beta
-\df{1}{2}\(\na^{\gamma\d\beta}{W_{\gamma\alpha}}^\beta\)K_{\beta\d\beta}, \\
{{\cal W}}^{\d\alpha} &=
(\b\sigma^{bc}\epsilon)^{\d\gamma\d\beta}
W^{\d\alpha}{}_{\d\beta\d\gamma}M_{cb}
-\df{1}{2}\(\b\na_{\d\gamma}{W}^{\d\gamma\d\alpha}_{
\hphantom{\d\gamma\d\alpha}\d\beta} \)\b{S}^{\d\beta}
-\df{1}{2}\( \na^{\d\gamma\beta}W_{\d\gamma}^{
\hphantom{{\d\gamma}}\d\alpha\d\beta} \)K_{\beta\d\beta},
   \end{split}
\end{equation}
and 
the parentheses for indices mean
symmetrizations of spinor indices:
$\psi_{(\alpha}\chi_{\beta)}=\tfrac{1}{2}(\psi_{\alpha}\chi_{\beta}+\psi_{\beta}\chi_{\alpha})$.
$W_{\alpha\beta\gamma}$ are 
chiral primary superfields
with Weyl weight 3/2 and chiral weight 1,
 and their indices are totally symmetric.
Here, a primary superfield is a superfield
that is invariant under the $K_A$-transformations: $K_A W_{\alpha\beta\gamma}=0$.
In particular, $T_{\alpha \d\beta}{}^c$ are given by
\begin{equation}
 T_{\alpha\d\beta}{}^c=2i(\sigma^c)_{\alpha\d\beta},
\end{equation}
which is used to solve Bianchi identities in appendix \ref{sec:der}.
\subsection{Abelian tensor hierarchy in conformal superspace}

We introduce antisymmetric tensor gauge fields into
conformal superspace.
Antisymmetric tensor gauge fields are expressed 
in terms of $p$-form gauge fields.
$p$-form gauge fields are transformed under 
Abelian internal gauge transformations
using $(p-1)$-form parameter superfields.
In addition, $p$-form gauge fields are 
shifted using $p$-form parameter superfields.
This structure of gauge transformation of the tensors 
is called an Abelian tensor hierarchy.

We explain the hierarchy concretely.
The $p$-form ($p\geq -1$) 
gauge fields $C_{[p]}^{I_p}$ are defined by%
\footnote{$(-1)$-forms are defined to be zero as
in the ordinary differential geometry.}
\begin{equation}
 C_{[p]}^{I_p}
:=\df{1}{p!}dz^{M_1}\wed \cdots \wed dz^{M_p} C_{M_p...M_1}^{I_p}
=\df{1}{p!}E^{A_1}\wed \cdots \wed E^{A_p} C_{A_p...A_1}^{I_p}.
\end{equation}
Here, the indices 
$I_p$ denote the indices of internal space of $p$-form 
$V_p$, which are assumed to be real vector spaces.
$I_p$ run over $1,...,\dim V_p$.
We denote infinitesimal internal 
gauge transformations of $p$-forms as $\delta_T(\Lambda)$,
where $\Lambda$ is a set of real $p$-form parameter 
superfields
$\Lambda^{I_{p+1}}_{[p]}$: $\Lambda=(\Lambda^{I_1}_{[0]},...,\Lambda^{I_4}_{[3]})$.
The gauge transformation laws of $C_{[p]}^{I_p}$ are given by
\begin{equation}
 \delta_T(\Lambda) C_{[p]}^{I_p} 
= d\Lambda_{[p-1]}^{I_p} +(q^{(p)}\cdot \Lambda_{[p]})^{I_p},
\end{equation}
where
$q^{(p)}$ are matrices which map $V_{p+1}$ to $V_p$.
$(q^{(p)}\cdot \Lambda_{[p]})^{I_p}$ are given by explicitly
 \begin{equation}
  (q^{(p)}\cdot \Lambda_{[p]})^{I_p}
= (q^{(p)})^{I_p}_{I_{p+1}} \Lambda_{[p]}{}^{I_{p+1}}.
 \end{equation}
We define the $X_{\cal A'}$ transformation laws of 
$C_{[p]}^{I_p}$ as
\begin{equation}
 \delta_G(\xi^{\cal C'}X_{\cal C'})C^{I_p}_{M_p...M_1}=0.
\label{eq:scinv}
\end{equation}
Field strengths of $p$-form gauge fields are defined by
\begin{equation}
\begin{split}
  F^{I_p}_{[p+1]}
&:= d C_{[p]}^{I_p}-(q^{(p)}\cdot C_{[p+1]})^{I_p} 
\\
&\hph{:}
=\df{1}{p!}dz^{M_1}\wed \cdots \wed dz^{M_p}\wed dz^N \der_N C_{M_p...M_1}^{I_p}
-\df{1}{(p+1)!}dz^{M_1}\wed \cdots \wed dz^{M_{p+1}}
 (q^{(p)}\cdot
 C_{M_{p+1}...M_1})^{I_p}.
\end{split}
\end{equation}
They are transformed under the internal gauge transformations
\begin{equation}
 \delta_T(\Lambda_{[p]}) 
F_{[p+1]}^{I_p}=
-(q^{(p)}\cdot q^{(p+1)}\cdot\Lambda_{[p+1]})^{I_p}.
\end{equation}
The invariances of the field strengths require that 
\begin{equation}
 q^{(p-1)}\cdot q^{(p)}=0.
\label{eq:qnilp}
\end{equation}

The SUSY transformations and spacetime translations 
are redefined with respect to 
$\delta_T$ transformations of $C_{[p]}^{I_p}$.
The redefinitions are the same as the case that
the tensor hierarchy does not exist \cite{bib:BG2}.
$P_A$-transformations are redefined by 
\begin{equation}
 \delta_G(\xi^AP_A)
=\delta_{\text{GC}}(\xi^M)
-\delta_G(\xi^Mh_M{}^{\cal A'}X_{\cal A'})
-\delta_T(\Lambda(\xi)).
\end{equation}
Here, $ \Lambda(\xi)$ is defined by
\begin{equation}
 \Lambda(\xi)
=(\iota_\xi C^{I_1}_{[1]},...,\iota_\xi C^{I_4}_{[4]}),
\end{equation}
 and $\iota_\xi $ is a interior product
\begin{equation}
 \iota_\xi C^{I_p}_{[p]}
=\df{1}{(p-1)!}
dz^{M_1}\wed \cdots \wed dz^{M_{p-1}}
\xi^{M_p}C_{M_p...M_1}^{I_p}.
\end{equation}
In particular, the $P_A$-transformations of $C^{I_p}_{[p]}$ are 
given by
\begin{equation}
  \delta_G(\xi^AP_A)C^{I_p}_{[p]}
=\delta_{\text{GC}}(\xi^M)C^{I_p}_{[p]}
-\delta_T(\Lambda(\xi))C^{I_p}_{[p]}
=\iota_\xi F^{I_p}_{[p+1]}.
\end{equation}
The $P_A$-transformation laws of superfields which are 
invariant under $\delta_T$ transformations are not changed.
Note that we obtain SUSY transformations of $p$-form gauge fields 
if we choose $\xi^A=\xi^{\ul\alpha}$.

Field strengths obey the following Bianchi identities:
\begin{equation}
 dF_{[p+1]}^{I_p}=-(q^{(p)}\cdot F_{[p+2]})^{I_p}.
\label{eq:BI}
\end{equation}
The existence of the tensor hierarchy deforms 
the Bianchi identites:
The tensor hierarchy relates 
the exterior derivatives on the $(p+1)$-form field strengths
 to the $(p+2)$-form field strengths.
The Bianchi identities play an important role 
in the next section.

Explicitly, we denote the $p$-form gauge fields,
the field strengths and the Bianchi identities 
in table \ref{tab:gfb}.
\begin{table}[h!]
\[
\begin{array}{c|c|c|c}
\hline\hline
\text{form}&\text{gauge field}&\text{field strength}
 &\text{Bianchi identity}
\\
\hline
\text{4-form}&\Tspan{U^{I_4}}
&G^{I_4}=dU^{I_4}=0&
-
\\
\text{3-form}&
\Tspan{C^{I_3}}
&\Sigma^{I_3}=dC^{I_3}-(q^{(3)}\cdot U)^{I_3}
&d\Sigma^{I_3}=0
\\
\text{2-form}&
B^{I_2}&H^{I_2}=dB^{I_2}-(q^{(2)}\cdot C)^{I_2}
&dH=-(q^{(2)}\cdot \Sigma)^{I_2}
\\
\text{1-form}&A^{I_1}&F^{I_1}=dA^{I_1}-(q^{(1)}\cdot B)^{I_1}
& dF^{I_1}=-(q^{(1)}\cdot H)^{I_1}
\\
\text{0-form}&f^{I_0}&g^{I_0}=df^{I_0}-(q^{(0)}\cdot A)^{I_0}
& dg^{I_0}=-(q^{(0)}\cdot F)^{I_0}
\\
\text{$-1$-form}&0&\omega^{I_{-1}} = -(q^{(-1)}\cdot f)^{I_{-1}}
& d\omega^{I_{-1}}=-(q^{(-1)}\cdot g)^{I_{-1}}
\\
\hline\hline
 \end{array}  
\]
 \caption{
The $p$-forms, their corresponding field strengths and Bianchi identities.
}
\label{tab:gfb}
\end{table}
In  table \ref{tab:gfb}, 4-form gauge fields appear.
The bosonic component of gauge fields $U_{qpnm}^{I_4}$ exist
 in principle,
but the bosonic components of the 
field strengths are zero: $G^{I_4}_{rqpnm}=0$.
This is because $G^{I_4}_{rqpnm}$ is the fifth rank 
antisymmetric tensor, which must be zero in 4D.
Thus, we impose by hand that 
field strengths $G^{I_4}$ are equal to zero as in 
Ref.~\cite{bib:SS}.
There are also 0-form ``field strengths'' in principle,
but $(-1)$-form gauge field does not exist.
Thus 0-form field strengths are defined by $d\omega^{I_{-1}}=-(q^{(-1)}\cdot g)^{I_{-1}}$.

From the higher-dimensional view point \cite{bib:BBLR},
$V_p$ can be understood as spaces of 
differential forms on extra dimensions.
The matrices $q^{(p)}$ are understood as the exterior derivative
with respect to extra dimensions.

\section{Constraints and Bianchi identities}
\label{sec:cBi}
In this section, constraints on the field strengths are imposed
to construct irreducible 
superfields.
We solve the Bianchi identities under these constraints.
As a result, each of field strengths is expressed 
in terms of the corresponding gauge invariant 
superfields straightforwardly.

\subsection{Constraints}
In the previous section, we have defined the field strengths
 of the $p$-form gauge fields.
The field strengths have redundant degrees of freedom,
and we eliminate them by imposing constraints.
We will take the constraints 
as the same ones without the tensor 
hierarchy~\cite{{bib:BG2},{Muller:1985vga}}.
The constraints are explicitly given by  table \ref{tab:constr}.
\begin{table}[h]
\[
\begin{array}{c|c}
\hline\hline
\text{form}&\text{constraints}
\\
\hline
\text{4-form}&
\Tspan{ 
G^{I_4}_{EDCBA}=0
}
\\
\hline
\text{3-form}&
\Tspan{
  \Sigma^{I_3}_{\ul\delta\, \ul\gamma\,\ul\beta A}
=\Sigma^{I_3}_{\delta\d\gamma ba}=0
}
\\
\hline
\text{2-form}&
\Tspan{
 H^{I_2}_{\ul\gamma\,\ul\beta\,\ul\alpha}
=H^{I_2}_{\gamma\beta a}=H^{I_2}_{\d\gamma\d\beta a}=0,
\quad
H^{I_2}_{\gamma\d\beta a}
=+2i(\sigma_a)_{\gamma\d\beta}L^{I_2}
}
\\
\hline
\text{1-form}
&
\Tspan{
 F^{I_1}_{\ul\alpha\,\ul\beta}=0
}
\\
\hline
\text{0-form}& 
\Tspan{
g^{I_0}_\alpha =i \na_\alpha \Psi^{I_0}, 
\quad 
g^{I_0}_{\d\beta}=-i\b\na_{\d\beta} \Psi^{I_0},
\quad 
K_A \Psi^{I_0}=0
}
\\
\hline\hline
 \end{array} 
\]
\caption{The constraints on the field strengths.
}
\label{tab:constr} 
\end{table}
Here, $L^{I_2}$ and $\Psi^{I_0}$ are real superfields.
In addition, we have imposed that $\Psi^{I_0}$ are 
primary superfields in  table \ref{tab:constr}.

We solve the Bianchi identities under these constraints.
On the one hand, the field strengths obey the Bianchi identities.
On the other hand, we impose the constraints on the field strengths.
The consistency between Bianchi identities and the constraints 
leads to new relations of the field strengths.
Since the constraints are imposed 
Lorentz covariantly,
it is convenient to express the Bianchi identities~\eqref{eq:BI}
by flat indices rather than curved indices:
\begin{equation}
\begin{split}
&\df{1}{(p+1)!}E^{A_1}\wed \cdots \wed E^{A_{p+1}} \wed E^B 
\na_B F^{I_p}_{A_p...A_{p+1}}
+\df{1}{p!2!}E^{A_1}\wed \cdots E^{A_p}\wed E^{B}\wed E^C
T_{CB}{}^{A_{p+1}}F^{I_p}_{A_{p+1}...A_1}\\
&=-\df{1}{(p+2)!}E^{A_1}\wed \cdots \wed E^{A_{p+2}} 
(q^{(p)}\cdot F_{A_{p+2}...A_1})^{I_p}. 
\end{split}
\label{eq:covB}
\end{equation}
This equation follows from 
 Eq.~\eqref{eq:tor} and \eqref{eq:scinv}.
Equation \eqref{eq:tor} is used to express
the exterior derivative on the vielbein 1-form $dE^A$ 
in terms of the torsion 2-form.
The exterior derivatives on gauge fields are 
written by covariant derivatives on the field strengths
using Eq.~\eqref{eq:scinv}.

\subsection{Solutions to the Bianchi identities}

In this subsection we summarize the solutions to 
the Bianchi identities of Table \ref{tab:gfb}
under the constraints of Table \ref{tab:constr}.
The details of the derivations are discussed in appendix \ref{sec:der}.
Each of the field strengths is expressed by a single 
gauge invariant superfield 
$(Y^{I_3},L^{I_2},W_\alpha^{I_1},\Psi^{I_0})$.
We find the Weyl weights and chiral ones $(\Delta, w)$ 
of the gauge invariant superfields.
We also find that these gauge invariant field strengths are 
primary superfields ($\Psi^{I_0}$ are imposed to be primary as in 
table \ref{tab:constr}).
The weights and $K_A$-invariance play an important role in 
constructing superconformally invariant actions.

\subsubsection{3-form gauge fields}
\label{ssec:Y}
For 3-form gauge fields, 
all the components of the field strengths
 are expressed in terms of 
 chiral superfields $Y^{I_3}$ and their 
complex conjugates $\b{Y}^{I_3}$.
They appear in the 2-spinor/2-vector components as
\begin{equation}
\Sigma^{I_3\d\delta\d\gamma}{}_{ba}
=
\df{1}{2}(\b\sigma_{ba}\epsilon)^{\d\delta\d\gamma} Y^{I_3},
\quad
\Sigma^{I_3}_{\delta\gamma ba}
=\df{1}{2}(\sigma_{ba}\epsilon)_{\delta\gamma} \b{Y}^{I_3}.
\label{eq:Y}
\end{equation}
From Eqs.~\eqref{eq:3confD}, \eqref{eq:3confA},
\eqref{eq:3confS} and
\eqref{eq:3confbS},
 $Y^{I_3}$ obey
\begin{equation}
 D Y^{I_3}=3Y^{I_3}, \quad AY^{I_3}=2i Y^{I_3}, \quad K_A Y^{I_3}=0.
\label{eq:Yconf}
\end{equation}
They mean that $Y^{I_3}$ are  primary 
superfields and the weights are
\begin{equation}
(\Delta, w) = (3,2).
\end{equation}
Similarly,   $\b{Y}^{I_3}$ obey
\begin{equation}
D \b{Y}^{I_3}=3 \b{Y}^{I_3}, \quad A \b{Y}^{I_3}=-2i  \b{Y}^{I_3}
, \quad K_A  \b{Y}^{I_3}=0.
\end{equation}
Other Bianchi identities lead to 
\begin{equation}
 \na_{\alpha}\b{Y}^{I_3}=0,
\quad
 \b\na_{\d\alpha}Y^{I_3}=0,
\label{eq:Ychi}
\end{equation}
which mean that 
$Y^{I_3}$ and $\b{Y}^{I_3}$ are chiral and anti-chiral
 superfields, respectively.
Furthermore, 
1-spinor/3-vector components are expressed in terms of 
spinor derivatives of $Y^{I_3}$ and their conjugates:
\begin{equation}
 \Sigma^{I_3\d\delta}{}_{cba}=+\df{1}{16}
\b\sigma^{d\d\delta\delta}
\epsilon_{dcba}\na_\delta Y^{I_3},
\quad
 \Sigma^{I_3}_{\delta cba}=-\df{1}{16}(\sigma^d)_{\delta\d\delta}
\epsilon_{dcba}\b\na^{\d\delta} \b{Y}^{I_3}.
\label{eq:fYbis}
\end{equation}
$\Sigma^{I_3}_{dcba}$ are identified as the 
imaginary parts of $\na^2 Y^{I_3}$:
\begin{equation}
 \Sigma^{I_3}_{ dcba}
=\df{i}{64}\epsilon_{dcba} (\na^2 Y^{I_3}-\b\na^2\b{Y}^{I_3}).
\label{eq:imFY}
\end{equation}
 We can understand the non-dynamical 
4-form field strength in terms of the $F$-component of $Y^{I_3}$
by the $\theta=\b\theta=0$ projection of both hand sides of 
this equation, where the $\theta=\b\theta=0$ projection is 
the projection from the superspace to the bosonic spacetime.
The solutions to the Bianchi identities for 3-form gauge fields 
are the same as the case without tensor hierarchy
\cite{{Binetruy:1996xw},{bib:BG2}}.
Note that  our normalization of
 $Y^{I_3}$ is equivalent to $8G^S$ in Ref.~\cite{bib:BBLR}.

\subsubsection{2-form gauge fields}
\label{ssec:L}
The field strengths of 2-form gauge fields are expressed 
by real superfields $L^{I_2}$.
We list the solutions to the Bianchi identities.
\begin{itemize}
 \item 1-spinor/2-vector components
\begin{equation}
 H^{I_2}_{\delta ba}
=2(\sigma_{ba})_\delta{}^\beta\na_\beta L^{I_2},
\quad
 H^{I_2\d\delta} {}_{ba}
=2(\b\sigma_{ba})^{\d\delta}{}_{\d\phi}\b\na^{\d\phi}L^{I_2},
\label{eq:fLbis}
\end{equation}
\item 3-vector components
\begin{equation}
 H^{I_2}_{fba}
=\df{1}{4}\epsilon_{fbag}(\b\sigma^g)^{\d\epsilon\epsilon}
[\na_\epsilon,\b\na_{\d\epsilon}]L^{I_2}.
\label{eq:vL} 
\end{equation}
\item Deformed linearity conditions 
\begin{equation}
 \na^2L^{I_2}
=\df{1}{4}(q^{(2)}\cdot \b{Y})^{I_2},
\quad
\b\na^2 L^{I_2}
=\df{1}{4}
(q^{(2)}\cdot Y)^{I_2}.
\label{eq:modL}
\end{equation}
\item $D$-, $A$-, $K_A$-transformation laws
\begin{equation}
 D L^{I_2}=2L^{I_2}, \quad
A L^{I_2}=0,\quad K_A L^{I_2}=0.
\end{equation}
Hence, we find
\begin{equation}
(\Delta, w) = (2,0).
\end{equation}
\end{itemize}
Note that the tensor hierarchy deforms ordinary linearity conditions
of $L^{I_2}$ by $q^{(2)}$.
\footnote{A deformed linear multiplet  in 4D ${\cal N}=1$
SUGRA is discussed in Ref.~\cite{Antoniadis:2014hfa}.}
If the tensor hierarchy does not exist,
the deformed linearity conditions reduce to the ordinary 
linearity conditions $\na^2L^{I_2}=\b\na^2L^{I_2}=0$.
Note that  our normalization of
$L^{I_2}$ is equivalent to $\tfrac{1}{2}H^M$ 
in Ref.~\cite{bib:BBLR}.
\subsubsection{1-form gauge fields}
\label{ssec:W}
The solutions to 1-form gauge fields are mostly the same as 
an ordinary Abelian case.
The field strengths are
 expressed in terms of the gaugino superfields $W_\alpha^{I_1}$
 and their conjugates.
The explicit solutions to the Bianchi identities are 
as follows.
\begin{itemize}
 \item 1-spinor/2-vector components
\begin{equation}
 F_{\d\beta,\alpha\d\alpha}^{I_1}
=-2\epsilon_{\d\beta\d\alpha} W_{\alpha}^{I_1},
\quad
F^{I_1}_{\beta,\alpha\d\alpha}
=-2\epsilon_{\beta\alpha}
\b{W}^{I_1}_{\d\alpha}.
\end{equation}
\item Chirality conditions 
\begin{equation}
\b\na_{\d\beta}W^{I_1}_\alpha=0,\quad
\na_\alpha \b{W}^{I_1}_{\d\beta}=0.
\end{equation}
\item 2-vector components
\begin{equation}
F^{I_1}_{ba}
=-\df{i}{2}
\(
(\sigma_{ba})_\beta{}^\alpha \na^\beta W^{I_1}_\alpha
-(\b\sigma_{ba})^{\d\beta}{}_{\d\alpha}
\b\na_{\d\beta}\b{W}^{I_1\d\alpha}
\).
\label{eq:1fs}
\end{equation}
\item Deformed reality conditions
\begin{equation}
 \na^\alpha W^{I_1}_\alpha-\b\na_{\d\alpha}\b{W}^{I_1 \d\alpha}
=-4i(q^{(1)}\cdot L)^{I_1}.
\label{eq:DW}
\end{equation}
\item $D$-, $A$-, $K_A$-transformation laws
\begin{equation}
\begin{split}
&
 D W_{\alpha}^{I_1}=\df{3}{2}W_{\alpha}^{I_1},
\quad
A W_{\alpha}^{I_1} =i W_{\alpha}^{I_1},
\quad
K_A W_{\alpha}^{I_1}=0,
\\
&
 D \b{W}^{I_1\d\alpha}=\df{3}{2}\b{W}^{I_1\d\alpha},
\quad
A \b{W}^{I_1\d\alpha} =-i \b{W}^{I_1\d\alpha},
\quad
K_A \b{W}^{I_1\d\alpha}=0.  
\end{split}
\end{equation}
Then, we find the weights of $W_{\alpha}^{I_1}$:
\begin{equation}
(\Delta, w) = (3/2,1).
\end{equation}
\end{itemize}
The reality conditions of 
$W_{\alpha}^{I_1}$ 
are deformed by the tensor hierarchy, i.e., $q^{(1)}$:
$L^{I_2}$ appear in the imaginary parts of
 $\na^\alpha W^{I_1}_{\alpha}$ in the presence of 
the tensor hierarchy.
The deformed reality conditions reduce to the ordinary 
reality conditions
$\na^\alpha W_\alpha^{I_1}=\b\na_{\d\alpha}\b{W}^{I_1\d\alpha}$
 if the tensor hierarchy does not exist.
Note that  our normalization of $W_\alpha^{I_1}$ is
 equivalent to that of Ref.~\cite{bib:BBLR}.

\subsubsection{0-form gauge fields}
\label{ssec:Psi}
The field strengths of 0-form gauge fields are expressed 
in terms of real primary superfields $\Psi^{I_0}$.
The solutions to the Bianchi identities are as follows.
\begin{itemize}
 \item Vector components
\begin{equation}
 g_{a}^{I_0}
=\df{1}{4i}(\b\sigma_a)^{\d\alpha\beta}
(\na_\beta g_{\d\alpha}^{I_0}+\b\na_{\d\alpha}
g_{\beta}^{I_0})
=-\df{1}{4} (\b\sigma_a)^{\d\alpha\beta}
[\na_\beta, \b\na_{\d\alpha}]\Psi^{I_0}.
\label{eq:0B1}
\end{equation}
\item Modified higher component conditions
\begin{equation}
\df{1}{4} \b\na^2\na_\alpha\Psi^{I_0}
=(q^{(0)}\cdot W_\alpha)^{I_0},
\quad
 \df{1}{4}\na^2\b\na_{\d\alpha} \Psi^{I_0}
=(q^{(0)}\cdot \b{W}_{\d\alpha})^{I_0}.
\label{eq:0B2}
\end{equation}
\item $D$-, $A$-transformation laws
\begin{equation}
 D\Psi^{I_0}=0,\quad A\Psi^{I_0}=0.
\end{equation}
We find the weights of  $\Psi^{I_0}$:
\begin{equation}
(\Delta, w) = (0,0).
\end{equation}
\end{itemize}
The conditions in 
Eq.~\eqref{eq:0B2} are a bit peculiar 
in the presence of tensor hierarchy $q^{(0)}$:
In the case of the absence of the tensor hierarchy, 
the Bianchi identities lead to 
the constraints
$\b\na^2\na_\alpha \Psi^{I_0}=0$ and
$\na^2\b\na_{\d\alpha} \Psi^{I_0}=0$. 
We can find  the expression of $\Psi^{I_0}$ 
which can be consistent with the constraints.
We can use chiral and anti-chiral primary superfields. 
Chiral primary 
superfields ${S}^{I_0}$ 
as well as anti-chiral primary superfields  $\b{S}^{I_0}$ 
satisfy 
$\b\na^2\na_\alpha{S}^{I_0} =\b\na^2\na_\alpha\b{S}^{I_0}=0$.
The field strengths of 0-form gauge fields $\Psi^{I_0}$
can be defined as the imaginary part of 
the chiral superfields:
\begin{equation}
 \Psi^{I_0}
=\df{1}{2i}({S}^{I_0}-\b{S}^{I_0}),
 \end{equation}
which are consistent with the constraints and
the solutions to Bianchi identities.
Note that ${S}^{I_0}$ can be understood as 
the prepotentials for the 0-form gauge fields.
Now we consider the case of the existence of 
the tensor hierarchy. 
$\Psi^{I_0}$ are deformed to 
\begin{equation}
 \Psi^{I_0}
=\df{1}{2i}({S}^{I_0}-\b{S}^{I_0})- (q^{(0)}\cdot V)^{I_0},
 \end{equation}
where $V^{I_1}$ are the prepotentials for 1-form gauge fields.
Using 
$W^{I_1}_\alpha=-\frac{1}{4}\b\na^2\na_\alpha V^{I_1}$,
we obtain 
\begin{equation}
\frac{1}{4}\b\na^2\na_\alpha \Psi^{I_0}
= (q^{(0)}\cdot W_\alpha)^{I_0}.
 \end{equation}
The results are consistent with Eq.~\eqref{eq:0B2}.
Note that $S^{I_0}$ are not gauge invariant in the presence of tensor hierarchy.

\section{Component formalism}
\label{sec:cf}

In this section we
show the correspondence between the conformal superspace and 
superconformal tensor calculus~\cite{bib:KU} using the results in 
Ref.~\cite{bib:KYY}.
Superconformal tensor calculus 
is presumably the most practically useful formalism.
We focus on the correspondence of the superfields 
$Y^{I_3}$, $L^{I_2}$, $W^{I_1}_\alpha$ and  $\Psi^{I_0}$,
which characterize the corresponding field strengths.

\subsection{Components of the superfields 
$Y^{I_3}$, $L^{I_2}$, $W^{I_1}_\alpha$ and $\Psi^{I_0}$}
\newcommand{\rhp}{{\cal P}_\text{R}}
We express  
$Y^{I_3}$, $L^{I_2}$, $W^{I_1}_\alpha$ and $\Psi^{I_0}$
within the superconformal tensor calculus.
In the superconformal tensor calculus,
we denote the components of 
a general complex multiplet $V_\Gamma$ with arbitrary Lorentz 
indices $\Gamma$ as
\begin{equation}
V_\Gamma = [C_\Gamma,Z_\Gamma,H_\Gamma,K_\Gamma,B_{a \Gamma}, 
\Lambda_\Gamma, D_\Gamma].
\end{equation}
The components of $V_\Gamma$ are expressed 
by corresponding primary superfields $\Phi_\Gamma$ 
as in table \ref{tab:KUB}.
In this table,
the symbol of ``$|$'' means the $\theta=\b\theta=0$ projection.
As already appeared in Sec.~\ref{sec:cBi}, 
the $\theta=\b\theta=0$ projection
is the projection from the 
superspace to the bosonic spacetime.
Component fields are obtained by 
$\theta=\b\theta=0$ projections of superfields.
\begin{table}[h]
\[
\begin{array}{c|l}
\hline\hline
\text{component}&\multicolumn{1}{c}{\text{superspace}}\\
\hline
 {C}_\Gamma &
\ \Tspan{\Phi_\Gamma|}
\\ \hline
 {Z}_\Gamma &
\Tspan{
\bmx
-i\na_\alpha \Phi_\Gamma \\
+i{\b\na}^{\d\alpha}\Phi_\Gamma
\emx \!\Big| }  
\\ \hline
 {H}_\Gamma &
\Tspan{ +\frac{1}{4}(\na^2 \Phi_\Gamma + \b\na^2 \Phi_\Gamma)|}
\\\hline
 {K}_\Gamma & 
\Tspan{-\frac{i}{4}(\na^2 \Phi_\Gamma - \b\na^2 \Phi_\Gamma)|}
\\ \hline
 {B}_{a\Gamma} &
\Tspan{-\frac{1}{4}(\b\sigma_a)^{\d\beta\beta}
[\na_\beta,{\b\na}_{\d\beta}]\Phi_\Gamma| } 
\\ \hline
 \Lambda_\Gamma & 
\Tspan{\df{i}{4}
\bmx
-\b\na^2\na_\alpha \Phi_\Gamma \\
+\na^2\b\na^{\d\alpha}\Phi_\Gamma
\emx \!\Big|
}
+2i
\bmx
{\cal W}_\alpha\\
{{\cal W}}^{\d\alpha}
\emx
\Phi_\Gamma|
\\
\hline
{D}_\Gamma&
\Tspan{\tfrac{1}{8}\b\na_{\d\alpha}\na^2\b\na^{\d\alpha}\Phi_\Gamma|
+{\cal W}_{\d\alpha}\b\na^{\d\alpha}\Phi_\Gamma | } 
\\  &
\qquad =
\Tspan{\tfrac{1}{8}\na^\alpha\b\na^2\na_\alpha \Phi_\Gamma|
-{\cal W}^\alpha \na_\alpha \Phi_\Gamma| }
\\
\hline
\hline
\end{array}
\]
\caption{The components of conformal multiplets.
The components are
expressed in terms of the $\theta=\b\theta=0$ projections of the
 superfields $\Phi_\Gamma$.
In this table,  
$[C_\Gamma,Z_\Gamma,H_\Gamma,K_\Gamma,B_{a \Gamma}, 
\Lambda_\Gamma, D_\Gamma]$
correspond to those of 
$[{\cal C}_\Gamma,{\cal Z}_\Gamma,{\cal H}_\Gamma,
{\cal K}_\Gamma,{\cal B}_{a \Gamma}, 
\Lambda_\Gamma, {\cal D}_\Gamma]$
 in Ref.~\cite{bib:KU}.}
\label{tab:KUB}
\end{table}

We also denote chiral conformal multiplets $T_\Gamma$ as
\begin{equation}
 T_\Gamma
=[A_\Gamma,\rhp \chi_\Gamma, F_\Gamma].
\end{equation}
$T_\Gamma$ are embedded into the general complex multiplets as
\begin{equation}
 V(T_\Gamma)
=[A_\Gamma,-i\rhp\chi_\Gamma, -F_\Gamma, iF_\Gamma,
 i\na_a A_\Gamma, 0,0].
\end{equation}
\newcommand{\lrb}{\left[ }
\newcommand{\rrb}{\right]}
\newcommand{\re}{\text{Re}\,}
\newcommand{\im}{\text{Im}\,}

The components of gauge invariant superfields $Y^{I_3}$, $L^{I_2}$,
$W_\alpha^{I_1}$ and $\Psi^{I_0}$ are given in tables \ref{tab:Y}, \ref{tab:L},
\ref{tab:W} and \ref{tab:Psi}, respectively.
In these tables, note that the tensor hierarchy deforms the higher components of 
the $L^{I_2}$, $W^{I_1}_\alpha$ and $\Psi^{I_0}$ in the presence of $q$'s.

\begin{table}[h]
\[
\begin{array}{c|c}
\hline\hline
\text{component } & \text{superfield}
\\
\hline
\Tspan{A(Y^{I_3})} & Y^{I_3}|
\\
\hline
\Tspan{\rhp\chi (Y^{I_3})} & \na_\alpha Y^{I_3}|
\\
\hline
F(Y^{I_3}) &
\Tspan{
 -\df{1}{8}(\na^2 Y^{I_3}+\b\na^2\b{Y}^{I_3})|
 -\df{i}{6}\epsilon^{dcba}\Sigma^{I_3}_{dcba}|
}
\\
\hline
\hline
\end{array} 
\]
\caption{The components of the chiral primary superfields 
$Y^{I_3}$. 
}
\label{tab:Y}
\end{table}

\begin{table}[h]
\[
\begin{array}{c|c}
\hline\hline
\text{component } & \text{superfield}
\\
\hline
\Tspan{C(L^{I_2})} & L^{I_2}|
\\
\hline
Z(L^{I_2}) & 
\Tspan{
\bmx
-i\na_\alpha L^{I_2}
\\
+i\b\na^{\d\alpha } L^{I_2}
\emx|
}
\\
\hline
H(L^{I_2}) &
\Tspan{
-\df{1}{16}(q^{(2)}\cdot (Y+\b{Y}))^{I_2}|
}
\\
\hline
K(L^{I_2}) &
\Tspan{
\df{i}{16}(q^{(2)}\cdot (Y-\b{Y}))^{I_2}|
}
\\
\hline
B_a(L^{I_2}) &
\Tspan{
-\df{1}{6}\epsilon_{adcb} H^{{I_2}dcb}|
}
\\
\hline
\Lambda(L^{I_2}) &
-i\stog{}{c}{\alpha}{\beta} \na_c 
\bmx
 -i\na_\beta L^{I_2}
\\
+i\b\na^{\d\beta} L^{I_2}
\emx|
+\Tspan{
\df{1}{16}\(q^{(2)}\cdot 
\bmx -i \na_\alpha Y
\\
+i\b\na^{\d\alpha} \b{Y}
\emx\)^{I_2}|
}
\\
\hline
D(L^{I_2}) &
-\na^a \na_a L^{I_2}|
+
\Tspan{
\df{1}{16}
\(
q^{(2)}\cdot \df{1}{4}(\na^2Y+\b\na^2\b{Y})
\)^{I_2}|
}
\\
\hline
\hline
\end{array} 
\]
\caption{
The components of the 
real primary superfields $L^{I_2}$.
}
\label{tab:L}
\end{table}
\clearpage

\begin{table}[h]
\[
\begin{array}{c|c}
\hline\hline
\text{component } & \text{superfield}
\\
\hline
\Tspan{A(W_\alpha^{I_1})} &
W^{I_1}_\alpha |
\\
\hline
\rhp \chi(W^{I_1}_\alpha) &
\Tspan{
\df{i}{2}(\sigma^{ba}\epsilon)_{\beta\alpha}F_{ba}^{I_1}|
+\df{1}{4}\epsilon_{\beta\alpha}
(\na^\gamma W^{I_1}_{\gamma}+\b\na_{\d\gamma}\b{W}^{I_1\d\gamma})|
-i \epsilon_{\beta\alpha}(q^{(1)}\cdot L)^{I_1}|
}
\\
\hline
F(W^{I_1}_\alpha) &
\Tspan{
-i \na_{\alpha\d\beta}\b{W}^{I_1 \d\beta}|
-2i (q^{(1)}\cdot L)^{I_1}|
}
\\
\hline
\hline
\end{array} 
\]
\caption{
The components of 
the chiral primary superfields $W_{\alpha}^{I_1}$.
The spinor index $\alpha$ is used for 
the external Lorentz index of $W^{I_1}_\alpha$.
}
\label{tab:W}
\end{table}

\begin{table}[h]
\[
\begin{array}{c|c}
\hline\hline
\text{component } & \text{superfield}
\\
\hline
C(\Psi^{I_0}) &
\Psi^{I_0} |
\\
\hline
Z(\Psi^{I_0}) &
\Tspan{
\bmx
-i\na_\alpha\Psi^{I_0}
\\
+i\b\na^{\d\alpha}\Psi^{I_0}
\emx
}|
\\
\hline
H(\Psi^{I_0}) &
\Tspan{
\df{1}{4}(\na^2\Psi^{I_0}+\b\na^2\Psi^{I_0})|
}
\\
\hline
K(\Psi^{I_0}) &
\Tspan{
-\df{i}{4}(\na^2\Psi^{I_0}-\b\na^2\Psi^{I_0})|
}
\\
\hline
B_a(\Psi^{I_0}) &
g^{I_0}_a|=\Tspan{-\df{1}{4}(\sigma_a)^{\d\alpha
\alpha}[\na_\alpha,\b\na_{\d\alpha}] \Psi^{I_0}|}
\\
\hline
\Lambda(\Psi^{I_0}) &
\Tspan{
\(
q^{(0)}\cdot
\bmx
-iW_\alpha
\\
+i\b{W}^{\d\alpha}
\emx
\)^{I_0}
|
}
\\
\hline
D(\Psi^{I_0}) &
\Tspan{
\df{1}{4}
\(q^{(0)}\cdot
(\na^\alpha W_\alpha+\b\na_{\d\alpha} \b{W}^{\d\alpha})
\)^{I_0}
|
}
\\
\hline
\hline
\end{array}  
\]
\caption{
The components of real primary superfields $\Psi^{I_0}$.
}
\label{tab:Psi}
\end{table}
\clearpage

\subsection{Bosonic field strengths}

In the previous subsection, 
the lowest component of the field strengths
$g^{I_0}_a|$, $F^{I_1}_{ab}|$, $H^{I_2}_{abc}|$ and 
$\Sigma_{abcd}^{I_3}|$ appear.
They are covariantly transformed under SUSY transformations,
because they have only Lorentz indices.
The lowest components of them
 are related to the lowest components of bosonic $p$-form 
gauge fields $C_{m_p...m_1}^{I_p}|$.
In SUGRA, they are also related to 
vierbein $e_m{}^a$ and gravitino $\psi_m{}^{\ul\alpha}$.
Thus, we express the lowest components of field strengths
in term of the lowest components of the bosonic $p$-form 
gauge fields,
vierbein, and gravitino. 

The expressions are obtained by
the so-called ``double bar projection''
\cite{{bib:BG2},{Baulieu:1986dp}}.
The double bar projections of the $p$-form gauge fields are 
defined as
\begin{equation}
 \df{1}{p!}dz^{M_1}\wed \cdots \wed dz^{M_p} C_{M_p...M_1}^{I_p}||
:=\df{1}{p!}dx^{m_1}\wed \cdots \wed dx^{m_p} C_{m_p...m_1}^{I_p}|.
\end{equation}
The symbol of ``$||$'' means the projection from superforms to 
forms on the bosonic spacetime:
ction: 
$d\theta^{\ul\mu}= \theta^{\ul\mu}= 0$.
However, there still exist fermion parts through contractions between indices
as shown below.

Explicitly, the double bar projections 
of the $p$-form gauge fields are as follows:
\begin{equation}
\begin{split}
 f^{I_0}||&=f^{I_0}|,
\\
 A^{I_1}||&=dx^m A^{I_1}_m|,
\\
B^{I_2}||&=\df{1}{2}dx^m\wed dx^n B_{nm}^{I_2}|,
\\
C^{I_3}||&=\df{1}{3!}dx^{m}\wed dx^n\wed dx^p C^{I_3}_{pnm}|,
\\
U^{I_4}||&=\df{1}{4!} dx^{m}\wed dx^n\wed dx^p\wed dx^q U^{I_4}_{qpnm}|.
\end{split}
\end{equation}
The double bar projections of the field strengths
systematically 
lead to the expressions of 
the bosonic field strengths.
We consider the simplest case.
For the 1-form field strength of 0-form gauge field,
the double bar projections of $g^{I_0}$ are
\begin{equation}
 g^{I_0}||
=dx^mg^{I_0}_m|=dx^mE_m{}^A g^{I_0}_A|
=dx^mE_m{}^a g^{I_0}_a|
+dx^mE_m{}^{\ul\alpha} g^{I_0}_{\ul\alpha}|.
\end{equation}
This relation gives rise to the component expression
\begin{equation}
\begin{split}
 g^{I_0}_a|
&=e_a{}^m
g^{I_0}_m|
-\df{1}{2}e_a{}^m\psi_m{}^\alpha g^{I_0}_\alpha|
-\df{1}{2}e_a{}^m\b\psi_{m\d\alpha} g^{I_0 \d\alpha}| 
\\
&
=
e_a{}^m (\der_m f^{I_0}-(q^{(0)}\cdot A_m)^{I_0})|
-\df{i}{2}e_a{}^m\psi_m{}^\alpha \na_\alpha \Psi^{I_0}|
+\df{i}{2}e_a{}^m\b\psi_{m\d\alpha} \b\na^{\d\alpha}\Psi^{I_0}|. 
\end{split}
\end{equation}
Here, we used
\begin{equation}
 dz^M g_M^{I_0}||=df^{I_0}||-(q^{(0)}\cdot A||)^{I_0}
=dx^m(\der_m f^{I_0}|-(q^{(0)}\cdot A_m|)^{I_0}).
\end{equation}
The same procedure can be applied to higher forms.
The results are summarized in appendix \ref{sec:bfs}.

\section{A superconformally invariant action}
\label{sec:sca}

In this section, we present a superconformally  invariant action
 which contains the kinetic terms of the 
$p$-form gauge fields.
In contrast to global SUSY case,
superconformally invariant actions require the 
conditions for Weyl and chiral weights $(\Delta,w)$ 
of integrands.
The integrands of F- and D-type actions must have the weights 
$(\Delta, w)=(3,2)$ and $(\Delta,w)=(2,0)$, respectively.
We compensate the weights when we construct invariant 
actions from $Y^{I_3}$, $L^{I_2}$, $W^{I_1}_\alpha$ and $\Psi^{I_0}$.
Such a procedure can be
 done by a so-called chiral compensator superfield $\Phi^\co$.
Chiral compensator is a chiral primary superfield with the weights $(\Delta,w)=(1,2/3)$.
The compensations are needed for $Y^{I_3}$ and $L^{I_2}$. 
We can rescale the weights of the superfields to
 $(\Delta,w)=(0,0)$ as
follows. 
\begin{equation}
 Y^{I_3}\quad \to \quad y^{I_3}:=\df{Y^{I_3}}{(\Phi^\co)^3},
\end{equation}
\begin{equation}
 L^{I_2} \quad \to\quad l^{I_2}:=\df{L^{I_2}}{\Phi^\co\b\Phi^\co}.
\end{equation}
Then $y^{I_2}$ and $l^{I_2}$ are weight $(0,0)$ primary superfields.
Recall that the former are chiral superfields and the latter are the real ones.
The compensations agree with those of \cite{{Dudas:2014pva}},
in which the tensor hierarchy does not exist. 
An invariant action is given by
\begin{equation}
S=-\df{3}{2}\int d^4x d^4\theta E \Phi^\co \b\Phi^\co e^{-K/3}
+\df{1}{4}\int d^4x d^2\theta {\cal E} 
g_{I_1J_1} W^{I_1\alpha}W_{\alpha}^{J_1}
+\int d^4x d^2\theta {\cal E}(\Phi^\co)^3W
+\text{h.c.}
\end{equation}
Here,
\begin{equation}
K=K(\Psi^{I_0},l^{I_2},y^{I_3}), \qquad W=W(y^{I_3}), \qquad
g_{I_1 J_1}=g(y^{I_3})_{I_1J_1}
\end{equation}
are the 
kinetic potential (rather than the K\"ahler potential) \cite{bib:BG2},
 the superpotential, and the gauge kinetic function respectively. All functions have the weights of $(\Delta,w)=(0,0)$.
Note that $K$ is a primary real superfield,
both $W$ and $g_{I_1 J_1}$ 
are primary chiral superfields. Further,
$K$ and $W$ are gauge invariant, whereas
$g_{I_1 J_1}$ is a function such that $g_{I_1 J_1} W^{I_1\alpha} W^{I_1}_\alpha$
is gauge invariant.
Such actions for tensors were also discussed in Ref.~\cite{bib:BG2}.
For example, in the case of quadratic kinetic terms,
$K$ is given as
\begin{equation}
 K=g_{I_0 J_0} \Psi^{I_0}\Psi^{J_0}
+g_{I_2J_2} l^{I_2}l^{J_2}
+g_{I_3J_3} y^{I_3}\b{y}^{J_3},
\end{equation}
where $g_{I_0 J_0}$, $g_{I_2 J_2}$ and $g_{I_3 J_3}$ are 
real constants. We may Taylor-expand $W$ as
\begin{equation}
W = \sum_n \sum_{\, \{I_3\}}
\lambda_{I_{3(1)} I_{3(2)}\cdots I_{3(n)}}  y^{I_{3(1)}} y^{I_{3(2)}} \cdots y^{I_{3(n)}},
\end{equation}
where $\lambda_{I_{3(1)} I_{3(2)}\cdots I_{3(n)}}$ are complex constants.
The expansion of this action will be done elsewhere 
\cite{bib:AHYY}.

We can reproduce the results in Ref.~\cite{bib:BG2}
by imposing the superconformal gauge fixing conditions.
The conditions are the same as those of
 Ref.~\cite{{bib:B1},{bib:KUigc}}.
We impose the conditions for
the compensator $\Phi^\co$ and dilatation gauge fields $B_M$ 
as follows:
\begin{equation}
\text{$D$-, $A$- gauge: } \Phi^\co=e^{K/6},\quad
\text{$K_A$- gauge: } B_M=0.
\end{equation} 

\section{Conclusions}
\label{sec:concl}

In this paper, we have considered a way to couple the 
Abelian tensor hierarchy to 4D ${\cal N}=1$ 
conformal supergravity.
We have used the conformal superspace formalism and
covariant approach.
The constraints on the field strengths 
have been imposed.
The constraints are the same as the case that
Abelian tensor hierarchy does not exist.
We have solved the 
Bianchi identities under the constraints.
Each of the field strengths is expressed in terms of 
the single superfield and its conjugate.
The linearity conditions which appear 
in 2-form gauge fields are deformed by the tensor hierarchy.
The reality conditions which appear in 1-form gauge fields 
are also deformed.
Furthermore, we have obtained nontrivial conditions 
of superconformal transformation laws.
We have also presented a superconformally invariant action.

There are remaining issues. 
The action of tensor superfields
in terms of their components should be considered. 
Such an action would be needed for
phenomenological applications. 
One can think also the Chern-Simons couplings of tensor fields. 
To introduce these terms, we need to
reconstruct the tensor hierarchy with the so-called prepotential
approach. 
Further, when there exist (non-Abelian ) gauge/matter fields,
one has to take chiral anomalies into account. Then the prepotential $S^{I_0}$ 
will appear also in the superpotential or gauge kinetic functions. 
We will address these issues elsewhere \cite{bib:AHYY}.

\subsection*{Acknowledgments}
This work is supported by Research Fellowships of Japan Society for
the Promotion of Science for Young Scientists Grant Numbers
16J06569 (S.A.), 16J03226 (R.Y.), and
JSPS KAKENHI Grant Number 26247042~(T.H.),
and MEXT-Supported Program for the Strategic Research
Foundation at Private Universities, ``Topological Science'', 
Grant Number S1511006 (T.H. and Y.Y.).
\appendix
\section{Notations}
\label{sec:not}
In this section, we summarize our notations.
The notations are the same as those of Wess-Bagger \cite{bib:WB}.
The Minkowski metric and the totally antisymmetric tensor are 
is given by
\begin{equation}
 \eta_{ab}=(-1,+1,+1,+1), 
\quad 
\epsilon^{0123}=-\epsilon_{0123}=+1.
\end{equation}
The standard contractions of two-component spinors are 
given by $\xi^\alpha\psi_\alpha$ and
$\b\xi_{\d\alpha}\b\psi^{\d\alpha}$.
The raising and lowering rules of indices are defined by
\begin{equation}
\psi^\alpha = \epsilon^{\alpha\beta}\psi_\beta,
\quad
\psi_\alpha=\epsilon_{\alpha\beta}\psi^\beta,
\quad
\b\psi_{\d\alpha}=\epsilon_{\d\alpha\d\beta}\b\psi^{\d\beta},
\quad
\b\psi^{\d\alpha}=\epsilon^{\d\alpha\d\beta}\b\psi_{\d\beta},
\end{equation}
where $\epsilon^{\alpha\beta}$,
$\epsilon_{\alpha\beta}$,
$\epsilon_{\d\alpha\d\beta}$
and
$\epsilon^{\d\alpha\d\beta}$
are antisymmetric tensors 
that satisfy
$\epsilon^{12}=\epsilon_{21}=+1$.
The Hermitian conjugate of a spinors is defined as
$(\psi_\alpha)^\dg=\b\psi_{\alpha}$.
Hermitian conjugate reverses the order of the product of spinors:
\begin{equation}
 (\psi_\alpha\xi_\beta)^\dg
=\b\xi_{\d\beta}\b\psi_{\d\alpha}.
\end{equation} 
Pauli matrices are defined as
\begin{equation}
 (\sigma_0)_{\alpha\d\beta}
=\bmx
1 & 0 \\ 0 & 1
\emx,
\quad
 (\sigma_1)_{\alpha\d\beta}
=\bmx
0 & 1 \\ 1 & 0
\emx,
\quad
 (\sigma_2)_{\alpha\d\beta}
=\bmx
0 & -i \\ i & 0
\emx,
\quad
 (\sigma_3)_{\alpha\d\beta}
=\bmx
1 & 0 \\ 0 & -1
\emx.
\end{equation}
Their Hermitian conjugates are given by
\begin{equation}
 (\b\sigma_a)^{\d\alpha\beta}
=(\sigma_a)^{\beta\d\alpha}
=\epsilon^{\d\alpha\d\gamma}\epsilon^{\beta\delta}
(\sigma_a)_{\delta\d\gamma}.
\end{equation}
Pauli matrices satisfy 
\begin{equation}
 (\sigma_a)_{\alpha\d\beta}(\b\sigma_b)^{\d\beta\gamma}
+ (\sigma_b)_{\alpha\d\beta}(\b\sigma_a)^{\d\beta\gamma}
=-2\eta_{ab}\delta_\alpha{}^\gamma,
\quad
 (\b\sigma_a)^{\d\alpha\beta}(\sigma_b)_{\beta\d\gamma}
+ (\b\sigma_b)^{\d\alpha\beta}(\sigma_a)_{\beta\d\gamma}
=-2\eta_{ab}\delta^{\d\alpha}{}_{\d\gamma},
 \end{equation}
\begin{equation}
(\sigma^a)_{\alpha\d\beta}
(\b\sigma_a)^{\d\gamma\delta}
=-2\delta_\alpha{}^\delta \delta^{\d\gamma}{}_{\d\beta}. 
\end{equation}
Lorentz vectors are expressed as mixed Lorentz spinors 
and vice versa:
\begin{equation}
 v_{\alpha\d\beta}=(\sigma^a)_{\alpha\d\beta}v_a,
\quad
 v_{a}=-\df{1}{2}(\b\sigma^a)^{\d\beta \alpha}v_{\alpha\d\beta},
\end{equation}
The matrices $\sigma_{ab}$ and $\b\sigma_{ab}$ are 
given by
\begin{equation}
 (\sigma_{ab})_{\alpha}{}^\beta
=\df{1}{4}((\sigma_a)_{\alpha\d\gamma}
(\b\sigma_b)^{\d\gamma\beta}
-(\sigma_b)_{\alpha\d\gamma}(\b\sigma_a)^{\d\gamma\beta}),
\quad
 (\b\sigma_{ab})^{\d\alpha}{}_{\d\beta}
=\df{1}{4}((\b\sigma_a)^{\d\alpha\gamma}
(\sigma_b)_{\gamma\d\beta}
-(\b\sigma_b)^{\d\alpha\gamma}(\sigma_a)_{\gamma\d\beta}).
\end{equation}
Any anti-symmetric tensor $F_{ab}$ can 
be decomposed into chiral and anti-chiral parts:
\begin{equation}
F_{ab} = -(\epsilon\sigma_{ab})^{\alpha\beta} F^-_{\alpha\beta}
+(\b{\sigma}_{ab}\epsilon)^{\d\alpha\d\beta} F^+_{\d\alpha\d\beta},  
\end{equation}
where 
\begin{equation}
F^-_{\alpha\beta} := \df{1}{2}(\sigma^{ab}\epsilon)_{\alpha\beta}F_{ab},
\qquad
F^+_{\d\alpha\d\beta} := -\df{1}{2}
(\epsilon\b\sigma^{ab})_{\d\alpha\d\beta}F_{ab}.
\label{eq:chiral_decom}
\end{equation}
In spinor notations, the equation is rewritten as
\begin{equation}
 F_{\alpha\d\alpha,\beta\d\beta}
=
(\sigma^a)_{\alpha\d\alpha}(\sigma^b)_{\beta\d\beta}
 F_{ab}
=-2 \epsilon_{\alpha\beta}F^+_{\d\alpha\d\beta}
+2\epsilon_{\d\alpha\d\beta}F^-_{\alpha\beta}.
\end{equation}
\section{Solving the Bianchi identities}
\label{sec:der}
In this appendix, we show detailed
derivations of solutions to the Bianchi identities.
Subsections \ref{ssec:Yd}, \ref{ssec:Ld}, \ref{ssec:Wd} and
\ref{ssec:Psid} are devoted to derive the results in 
\ref{ssec:Y}, \ref{ssec:L}, \ref{ssec:W} and
\ref{ssec:Psi}, respectively.
In the subsection \ref{ssec:sc}, we show derivations of the 
$D$-, $A$-, $K_A$-transformations of the 
gauge invariant superfields.
\subsection{The Bianchi identities for 3-form gauge fields}
\label{ssec:Yd}
Firstly, we solve the Bianchi identities for 3-form gauge fields.
The Bianchi identities are
given by
\begin{equation}
 \df{1}{4!}E^A\wed E^B \wed E^C\wed E^D \wed E^E\na_E 
\Sigma^{I_3}_{DCBA}
+\df{1}{3!2!} 
E^A\wed E^B \wed E^C\wed E^D \wed E^E T_{ED}{}^F \Sigma^{I_3}_{FCBA}
=0.
\end{equation}
Explicitly they are written by
\begin{equation}
\begin{split}
0&=
\hph{+}
\na_E\Sigma^{I_3}_{DCBA}
+
\na_D\Sigma^{I_3}_{CBAE}
+
\na_C\Sigma^{I_3}_{BAED}
+
\na_B\Sigma^{I_3}_{AEDC}
+
\na_A\Sigma^{I_3}_{EDCB}\\
&\quad+ 
T_{ED}{}^F\Sigma^{I_3}_{FCBA}
 - 
 T_{EC}{}^F\Sigma^{I_3}_{FDBA}
 - 
T_{EB}{}^F\Sigma^{I_3}_{FCDA}
-
T_{EA}{}^F\Sigma^{I_3}_{FCBD}
\\
&\quad
 + 
T_{DC}{}^F\Sigma^{I_3}_{FEBA}
-
T_{DB}{}^F\Sigma^{I_3}_{FECA}
-
T_{DA}{}^F\Sigma^{I_3}_{FEBC}
\\
&\quad
 + 
T_{CB}{}^F\Sigma^{I_3}_{FEDA}
-
T_{CA}{}^F\Sigma^{I_3}_{FBED}
\\
&\quad
-
T_{BA}{}^F\Sigma^{I_3}_{FECD}.
\end{split}
\label{eq:3B}
\end{equation}
The constraints for 3-form are given in table~\ref{tab:constr}.
Under these constraints, we solve the Bianchi identities.

For $E=\epsilon$, $D=\d\delta$, $C=\gamma$, $B=\beta$, $A=a$,
the Bianchi identities are
\begin{equation}
\begin{split}
0&=
\hph{+} 
T_{\epsilon\d\delta}{}^f\Sigma^{I_3}_{f\gamma\beta a}
+
T_{\d\delta\gamma}{}^f\Sigma^{I_3}_{f\epsilon\beta a}
 + 
T_{\d\delta\beta}{}^f\Sigma^{I_3}_{f\epsilon\gamma a}.
\end{split}
\end{equation}
This equation is equivalently written as
\begin{equation}
 \Sigma^{I_3}_{\epsilon\d\delta,\gamma\beta, \alpha\d\alpha}
+\Sigma^{I_3}_{\gamma\d\delta,\epsilon\beta,  \alpha\d\alpha}
=-\Sigma^{I_3}_{\beta\d\delta,\epsilon\gamma,  \alpha\d\alpha}.
\label{eq:3B1}
\end{equation}
We decompose $ \Sigma^{I_3}_{\epsilon\d\delta,\gamma,\beta, \alpha\d\alpha}
$ into 
chiral part and anti-chiral part as
\begin{equation}
 \Sigma^{I_3}_{\epsilon\d\delta,\gamma,\beta, \alpha\d\alpha}
=
-2
\epsilon_{\epsilon\alpha }
\Sigma^{+I_3}_{\d\delta\d\alpha,\gamma,\beta}
+
2
\epsilon_{\d\delta
\d\alpha}\Sigma^{-I_3}_{\epsilon\alpha,\gamma,\beta}.
\end{equation}
Substituting this decomposition into Eq.~\eqref{eq:3B1}, we obtain
\begin{equation}
 0=-2\epsilon_{\epsilon\alpha}
\Sigma^{+I_3}_{\d\delta\d\alpha,\gamma,\beta}
+
2\epsilon_{\d\delta\d\alpha}
\Sigma^{-I_3}_{\epsilon\alpha,\gamma,\beta}
-2
\epsilon_{\gamma\alpha}
\Sigma^{+I_3}_{\d\delta\d\alpha,\epsilon,\beta}
+
2
\epsilon_{\d\delta\d\alpha}
\Sigma^{-I_3}_{\gamma\alpha,\epsilon,\beta}
-2
\epsilon_{\beta\alpha}
\Sigma^{+I_3}_{\d\delta\d\alpha,\gamma,\epsilon}
+
2
\epsilon_{\d\delta\d\alpha}
\Sigma^{-I_3}_{\beta\alpha,\gamma,\epsilon}. 
\label{eq:3B1-1}
\end{equation}
 We find that anti-chiral parts vanish:
\begin{equation}
\Sigma^{+I_3}_{\d\delta\d\alpha,\gamma,\beta}=0. 
\end{equation}
Eq.~\eqref{eq:3B1-1} is then expressed as
\begin{equation}
 \Sigma^{-I_3}_{\gamma\alpha,\epsilon,\beta}
+\Sigma^{-I_3}_{\beta\alpha,\epsilon,\gamma}
+\Sigma^{-I_3}_{\epsilon\alpha,\gamma,\beta}=0,
\end{equation}
where we used 
$\Sigma^{-I_3}_{\gamma\alpha,\epsilon,\beta}
=\Sigma^{-I_3}_{\gamma\alpha,\beta,\epsilon}$.
Contracting $\gamma$ and $\alpha$ by $\epsilon^{\gamma\alpha}$, 
we obtain
\begin{equation}
0=\Sigma^{-I_3}{}_{\beta}{}^\gamma{}_{,\epsilon,\gamma}
+\Sigma^{-I_3}{}_{\epsilon}{}^\gamma{}_{,\gamma,\beta}.
\end{equation}
This equation means that
\begin{equation}
 \Sigma^{-I_3}{}_{\beta}{}^\gamma{}_{,\epsilon,\gamma}
=\df{1}{2}\epsilon_{\beta\epsilon}
\Sigma^{-I_3\delta\gamma}{}_{,\delta,\gamma}.
\end{equation}
Then,  $\Sigma^{-I_3}_{\epsilon\alpha,\gamma,\beta}$
are calculated as
\begin{equation}
\begin{split}
\Sigma^{-I_3}_{\epsilon\alpha,\gamma,\beta}
&=
\df{1}{2}
(
\Sigma^{-I_3}_{\epsilon\alpha,\gamma,\beta}
+
\Sigma^{-I_3}_{\beta\alpha,\gamma,\epsilon}
)
+\df{1}{2}
\epsilon_{\epsilon\beta}
\Sigma^{-I_3}{}^\delta{}_{\alpha}{}_{,\gamma,\delta}
\\
&=-\df{1}{2}
\Sigma^{-I_3}_{\alpha\gamma,\epsilon,\beta}
+\df{1}{2}
\epsilon_{\epsilon\beta}
\Sigma^{-I_3}{}^\delta{}_{\alpha}{}_{,\gamma,\delta}
\\
&=-\df{1}{2}
\Sigma^{-I_3}_{\alpha\epsilon,\gamma,\beta}
-\df{1}{2}
(
\Sigma^{-I_3}_{\alpha\gamma,\epsilon,\beta}
-\Sigma^{-I_3}_{\alpha\epsilon,\gamma,\beta}
)
+\df{1}{2}
\epsilon_{\epsilon\beta}
\Sigma^{-I_3}{}^\delta{}_{\alpha}{}_{,\gamma,\delta}
\\
&=-\df{1}{2}
\Sigma^{-I_3}_{\alpha\epsilon,\gamma,\beta}
-\df{1}{2}
\epsilon_{\gamma\epsilon}
\Sigma^{-I_3}{}_{\alpha}{}^\delta{}_{,\delta,\beta}
+\df{1}{2}
\epsilon_{\epsilon\beta}
\Sigma^{-I_3}{}^\delta{}_{\alpha}{}_{,\gamma,\delta}
\\
&=-\df{1}{2}
\Sigma^{-I_3}_{\alpha\epsilon,\gamma,\beta}
-\df{1}{4}
\epsilon_{\gamma\epsilon}
\epsilon_{\alpha\beta}
\Sigma^{-I_3\zeta\delta}{}_{,\delta,\zeta}
+\df{1}{4}
\epsilon_{\epsilon\beta}\epsilon_{\alpha\gamma}
\Sigma^{-I_3\delta\zeta}{}_{,\zeta,\delta}.
\end{split}
\end{equation}
Therefore, 
$ \Sigma^{-I_3}_{\epsilon\alpha,\gamma,\beta}$ 
have only scalar components:
\begin{equation}
 \Sigma^{-I_3}_{\epsilon\alpha,\gamma,\beta}
 =
\df{1}{6}
(
\epsilon_{\alpha\beta}
\epsilon_{\epsilon\gamma}
+\epsilon_{\alpha\gamma}\epsilon_{\epsilon\beta}
)
\Sigma^{-I_3\eta\zeta}{}_{,\zeta,\eta}.
\end{equation}
Thus, we obtain
\begin{equation}
 \Sigma^{I_3}_{\epsilon\d\delta,\gamma,\beta, \alpha\d\alpha}
=
\df{1}{3}\epsilon_{\d\delta\d\alpha}
(
\epsilon_{\alpha\beta}
\epsilon_{\epsilon\gamma}
+\epsilon_{\alpha\gamma}\epsilon_{\epsilon\beta}
)
\Sigma^{-I_3\eta\zeta}{}_{,\zeta,\eta}.
\label{eq:bY0}
\end{equation}
We define $\b{Y}^{I_3}$ as 
\begin{equation}
 \b{Y}^{I_3}:= \df{2}{3}\Sigma^{-I_3\eta\zeta}{}_{,\zeta,\eta}.
\label{eq:3def1}
\end{equation}
This definitions of $\b{Y}^{I_3}$ agree with those of 
Eq.~(VI-2.7) in Ref.~\cite{bib:BG2}.
Equation~\eqref{eq:bY0} 
is equivalently expressed as in Eq.~\eqref{eq:Y}.
Similarly, dotted versions of the 
Bianchi identities \eqref{eq:3B1} lead
 to Eq.~\eqref{eq:Y}.

For $E=\epsilon$, $D=\delta$, $C=\gamma$, $B=\beta\d\beta$, 
$A=\alpha\d\alpha$, Eq.~\eqref{eq:3B} is written as
\begin{equation}
\begin{split}
 0&
=\na_\epsilon \Sigma^{I_3}_{\delta,\gamma,\beta\d\beta,\alpha\d\alpha}
+\na_\delta \Sigma^{I_3}_{\gamma,\beta\d\beta,\alpha\d\alpha,\epsilon}
+\na_\gamma \Sigma^{I_3}_{\beta\d\beta,\alpha\d\alpha,\epsilon,\delta}
\\
&=
\epsilon_{\d\beta\d\alpha}(
\epsilon_{\beta\delta}\epsilon_{\alpha\gamma}
+\epsilon_{\alpha\delta}\epsilon_{\beta\gamma}
)\na_{\epsilon}\b{Y}^{I_3}
+
\epsilon_{\d\beta\d\alpha}(
\epsilon_{\beta\gamma}\epsilon_{\alpha\epsilon}
+\epsilon_{\alpha\gamma}\epsilon_{\beta\epsilon}
)\na_{\delta}\b{Y}^{I_3}
+
\epsilon_{\d\beta\d\alpha}(
\epsilon_{\beta\epsilon}\epsilon_{\alpha\delta}
+\epsilon_{\alpha\epsilon}\epsilon_{\beta\delta}
)\na_{\gamma}\b{Y}^{I_3}.
\end{split}
\end{equation}
Contracting spinors by 
$\epsilon^{\d\alpha\d\beta}\epsilon^{\delta\beta}\epsilon^{\gamma\alpha}
$,
 we obtain the anti-chirality conditions of $\b{Y}^{I_3}$
 in Eq.~\eqref{eq:Ychi}.
Similarly, we obtain chirality conditions of $Y^{I_3}$ as in
Eq.~\eqref{eq:Ychi}. 

For $E=\epsilon$, $D=\d\delta$, $C=\d\gamma$ $B=\beta\d\beta$,
$A=\alpha\d\alpha$, Eq.~\eqref{eq:3B} is written as
\begin{equation}
0=\na_\epsilon
 \Sigma^{I_3}_{\d\delta,\d\gamma,\beta\d\beta,\alpha\d\alpha}
+T_{\epsilon\d\delta}{}^f
\Sigma^{I_3}_{f,\d\gamma,\beta\d\beta,\alpha\d\alpha}
+T_{\epsilon\d\gamma}{}^f
\Sigma^{I_3}_{f,\d\delta,\beta\d\beta,\alpha\d\alpha}.
\end{equation}
This is solved as
\begin{equation}
 \na_{\delta}
 Y^{I_3}=+\df{2}{3}\epsilon^{dcba}\sigma_{d\delta\d\delta}
\Sigma^{I_3\d\delta}{}_{cba},
\end{equation}
or equivalently expressed as in Eq.~\eqref{eq:fYbis}.
Similarly, $\Sigma^{I_3}_{\delta cba}$  are 
expressed in terms of $ \b\na^{\d\delta} \b{Y}^{I_3}$:
\begin{equation}
 \b\na^{\d\delta}
 \b{Y}^{I_3}=-\df{2}{3}\epsilon^{dcba}\b\sigma_d^{\d\delta\delta}
\Sigma^{I_3}_{\delta cba}, 
\label{eq:fYhc}
\end{equation}
or Eq.~\eqref{eq:fYbis}.

For $E=\epsilon$, $D=\d\delta$, $C=c$, $B=b$, $A=a$,
 Eq.~\eqref{eq:3B} is expressed as
\begin{equation}
 0=\na_\epsilon \Sigma^{I_3\d\delta}{}_{ cba}
-\b\na^{\d\delta} \Sigma^{I_3}_{cba\epsilon}
+T_{\epsilon}{}^{\d\delta f}\Sigma^{I_3}_{ fcba}.
\end{equation}
Using \eqref{eq:fYbis}, and contracting spinors, we
obtain 
\begin{equation}
 \df{8}{3}i\epsilon^{dcba}\Sigma^{I_3}_{ dcba}
=\na^2Y^{I_3}-\b\na^2\b{Y}^{I_3}.
\end{equation}
They are equivalently written as in Eq.~\eqref{eq:imFY}.
There is no more non-trivial Bianchi identity from constraints.

\subsection{The Bianchi identities for 2-form gauge fields}
\label{ssec:Ld}
Next, we solve the Bianchi identities for 2-form gauge fields.
The Bianchi identities are written as
\begin{equation}
\begin{split}
0&=
 \na_D H^{I_2}_{CBA}-\na_C H^{I_2}_{BAD}+\na_B H^{I_2}_{ADC}
-\na_AH^{I_2}_{DCB}\\
&\quad+T_{DC}{}^EH^{I_2}_{EBA}
-T_{DB}{}^EH^{I_2}_{ECA}
+T_{DA}{}^EH^{I_2}_{ECB}
-T_{CB}{}^EH^{I_2}_{EAD}
+T_{CA}{}^EH^{I_2}_{EBD}
+T_{BA}{}^EH^{I_2}_{EDC}\\
&\quad +(q^{(2)}\cdot \Sigma_{DCBA})^{I_2}.
\end{split}
\label{eq:2B}
\end{equation}
The constraints on the 
field strengths of 2-form gauge fields are given in 
table~\ref{tab:constr}.

For $D=\delta$, $C=\gamma$, $B=\d\beta$, $a=\alpha\d\alpha$,
Eq.~\eqref{eq:2B} is 
\begin{equation}
0= \na_\delta H^{I_2}_{\gamma,\d\beta,\alpha\d\alpha}
-
\na_\gamma H^{I_2}_{\d\beta,\alpha\d\alpha,\delta}
+
T_{\delta\d\beta}{}^e H^{I_2}_{e,\gamma,\alpha\d\alpha}
-
T_{\d\gamma\beta}{}^e H^{I_2}_{e,\alpha\d\alpha,\delta}.
\end{equation}
Using the constraints in table~\ref{tab:constr}, we obtain
\begin{equation}
-4i
\epsilon_{\gamma\alpha}\epsilon_{\d\beta\d\alpha}
 \na_\delta L^{I_2}
-4i
\epsilon_{\delta\alpha}\epsilon_{\d\beta\d\alpha}
\na_\gamma L^{I_2}
+2i
H^{I_2}_{\delta\d\beta,\gamma,\alpha\d\alpha}
-2i
H^{I_2}_{\gamma\d\beta,\alpha\d\alpha,\delta}=0.
\label{eq:2B2-1}
\end{equation}
We decompose $H^{I_2}_{\delta,\gamma\d\beta,\alpha\d\alpha}$ as
\begin{equation}
 H^{I_2}_{\delta,\gamma\d\beta,\alpha\d\alpha}
=-2\epsilon_{\gamma\alpha}H^{I_2 +}{}_{\delta,\d\beta\d\alpha}
+2\epsilon_{\d\beta\d\alpha}H^{I_2 -}{}_{\delta,\gamma\alpha}.
\label{eq:2B2-3}
\end{equation}
Substituting this into 
Eq.~\eqref{eq:2B2-1} and contracting spinors
 by $\epsilon^{\d\beta\d\alpha}$, we obtain
\begin{equation}
 \epsilon_{\alpha\gamma}\na_\delta L^{I_2}
+\epsilon_{\alpha\delta}\na_\gamma L^{I_2}
-
H^{I_2-}{}_{\gamma,\delta\alpha}
-
H^{I_2-}{}_{\delta,\gamma\alpha}
=0.
\label{eq:2B2-2}
\end{equation}
Furthermore, contracting spinor indices 
by $\epsilon^{\alpha\gamma}$, we find that 
\begin{equation}
H^{I_2 - \alpha,}{}_{\delta\alpha}
=-3\na_\delta L^{I_2}.
\end{equation}
Substituting this equation to Eq.~\eqref{eq:2B2-2},
we obtain
\begin{equation}
H^{I_2-}{}_{\alpha,\gamma\delta}
=
-\epsilon_{\alpha\gamma}\na_\delta L^{I_2}
-\epsilon_{\alpha\delta}\na_\gamma L^{I_2},
\end{equation}
where we used 
$H^{I_2-}{}_{\gamma,\delta\alpha}
=\epsilon_{\gamma\alpha}H^{I_2- \phi}{}_{,\delta\phi}
+H^{I_2-}{}_{\alpha,\delta\gamma}$ and
$H^{I_2-}{}_{\delta,\gamma\alpha}
=\epsilon_{\delta\alpha}H^{I_2- \phi}{}_{,\gamma\phi}
+H^{I_2-}{}_{\alpha,\gamma\delta}$.

The symmetrization of $\d\alpha\corr\d\beta$ 
in Eq.~\eqref{eq:2B2-1} with Eq.~\eqref{eq:2B2-3} reads
\begin{equation}
 H^{I_2+}{}_{\delta,\d\beta\d\alpha}=0.
\end{equation}
Therefore, we obtain
\begin{equation}
 H^{I_2}_{\delta,\gamma\d\beta,\alpha\d\alpha}
=-2\epsilon_{\d\beta\d\alpha}
(\epsilon_{\delta\gamma}\na_\alpha L^{I_2}
+\epsilon_{\delta\alpha}\na_\gamma L^{I_2}),
\label{eq:fL}
\end{equation}
which is equivalent to  Eq.~\eqref{eq:fLbis}.
Similarly, 
for $D=\d\delta$, $C=\d\gamma$, $B=\beta$, $a=\alpha\d\alpha$,
we obtain 
\begin{equation}
 H^{I_2}_{\d\gamma,\beta\d\delta,\alpha\d\alpha}
=-2\epsilon_{\beta\alpha}
(\epsilon_{\d\gamma\d\delta}\b\na_{\d\alpha}L^{I_2}
+
\epsilon_{\d\gamma\d\alpha}\b\na_{\d\delta}L^{I_2}
),
\end{equation}
which is equivalent to Eq.~\eqref{eq:fLbis}.

For $D=\delta$, $C=\d\gamma$, $B=b$, $A=a$, 
Eq.~\eqref{eq:2B} is written as
\begin{equation}
0= \na_\delta H^{I_2}_{\d\gamma ba}
+
\b\na_{\d\gamma}H^{I_2}_{ba\delta}
+\na_{b}H^{I_2}_{a\delta\d\gamma}
-
\na_{a}H^{I_2}_{\delta\d\gamma b}
+T_{\delta\d\gamma}{}^eH^{I_2}_{eba}.
\end{equation}
This equation implies
\begin{equation}
 0=2(\b\sigma_{ba})_{\d\gamma\d\phi}\na_\delta\b\na^{\d\phi}L^{I_2}
+2(\sigma_{ba})_\delta{}^\phi \b\na_{\d\gamma}\na_\phi L^{I_2}
+2i(\sigma_a)_{\delta\d\gamma}\na_b L^{I_2}
-2i(\sigma_b)_{\delta\d\gamma}\na_a L^{I_2}
+2i(\sigma_{e})_{\delta\d\gamma}H^{I_2}_{eba}.
\end{equation}
From this identity, we obtain
\begin{equation}
\epsilon^{cfba}(\sigma_c)_{\gamma\d\gamma}H^{I_2}_{fba}
=-3[\na_\gamma,\b\na_{\d\gamma}]L^{I_2},
\end{equation}
which is equivalent to Eq.~\eqref{eq:vL}.

For $D=\delta$, $C=\gamma$, $B=b$, $A=a$, 
Eq.~\eqref{eq:2B} is expressed as
\begin{equation}
0= \na_\delta H^{I_2}_{\gamma ba}
+
\na_\gamma H^{I_2}_{\delta ba}
+
(q^{(2)}\cdot \Sigma_{\delta\gamma ba})^{I_2}.
\end{equation}
Using Eqs.~\eqref{eq:Y} and \eqref{eq:fLbis},
we obtain Eq.~\eqref{eq:modL}.
Similarly, for $D=\d\delta$, $C=\d\gamma$, $B=b$, $A=a$, 
we find Eq.~\eqref{eq:modL}.

\subsection{The Bianchi identities for 1-form gauge fields}
\label{ssec:Wd}

Thirdly, we solve the Bianchi identities for the field strengths
of 1-form gauge fields \cite{bib:WB}:
\begin{equation}
0= \na_C F^{I_1}_{BA}+\na_B F^{I_1}_{AC}+\na_AF^{I_1}_{CB}
+T_{CB}{}^DF^{I_1}_{DA}+T_{BA}{}^DF^{I_1}_{DC}+T_{AC}{}^DF^{I_1}_{DB}
+(q^{(1)}\cdot H_{CBA})^{I_1}.
\label{eq:1B}
\end{equation}
For $C=\d\gamma$, $B=\beta$, $A=\alpha$, Bianchi identities are
\begin{equation}
0=T_{\d\gamma \beta}{}^d F^{I_1}_{d\alpha}
+
 T_{\alpha\d\gamma}{}^d F^{I_1}_{d\beta}. 
\end{equation}
This means symmetric part of undotted spinors in 
$F^{I_1}_{\d\gamma\beta,\alpha}$ is equal to zero.
Then, we can write 
\begin{equation}
 F^{I_1}_{\alpha,\beta\d\gamma}
=-2\epsilon_{\alpha\beta}W^{I_1}_{\d\gamma}.
\label{eq:1def1}
 \end{equation}
Similarly, for $C=\gamma$, $B=\d\beta$, $A=\d\alpha$,
we obtain
\begin{equation}
 F^{I_1}_{\d\alpha,\gamma\d\beta}
=-2\epsilon_{\d\alpha\d\beta}W^{I_1}_{\gamma}. 
\label{eq:1def2}
\end{equation}
For $C=\d\gamma$, $B=\d\beta$, $A=\alpha\d\alpha$,
the Bianchi identities are
\begin{equation}
0=
\b\na_{\d\gamma} F^{I_1}_{\d\beta,\alpha\d\alpha}
-
\b\na_{\d\beta} F^{I_1}_{\alpha\d\alpha,\d\gamma}.
\end{equation}
Using \eqref{eq:1def1} we obtain chirality condition for
$W^{I_1}_\alpha$ as
\begin{equation}
 \b\na_{\d\gamma}W^{I_1}_{\alpha}=0.
\end{equation}
Similarly for $C=\gamma$, $B=\beta$, $A=\alpha\d\alpha$, we find
\begin{equation}
 \na_\gamma W^{I_1}_{\d\alpha}=0.
\end{equation}
For $C=\gamma$, $B=\d\beta$, $A=\alpha\d\alpha$, 
the Bianchi identities are
\begin{equation}
0=
 \na_\gamma F^{I_1}_{\d\beta,\alpha\d\alpha}
-
\b\na_{\d\beta}F^{I_1}_{\alpha\d\alpha,\gamma}
+T_{\gamma\d\beta}{}^dF^{I_1}_{d,\alpha\d\alpha}
-4i
\epsilon_{\gamma\alpha}\epsilon_{\d\beta\d\alpha}
(q^{(1)}\cdot L)^{I_1}.
\label{eq:1B1}
\end{equation}
Contracting the spinor indices by
$\epsilon^{\alpha\gamma}\epsilon^{\d\alpha\d\beta}$, we obtain
Eq.~\eqref{eq:DW}.
Then, symmetrizing spinors in Eq.~\eqref{eq:1B1}, we 
also obtain Eq.~\eqref{eq:1fs}.

\subsection{The Bianchi identities for 0-form gauge fields}
\label{ssec:Psid}
Finally, we solve the Bianchi identities for 0-form gauge fields.
The Bianchi identities are given by 
\begin{equation}
0= \na_B g^{I_0}_A-\na_A g^{I_0}_B
+T_{BA}{}^C g^{I_0}_C
+(q^{(0)}\cdot F_{BA})^{I_0}.
\label{eq:0B}
\end{equation}
For $B=\beta$, $A=\alpha$, Eq.~\eqref{eq:0B} is 
\begin{equation}
0= \na_\beta g^{I_0}_\alpha + \na_\alpha g^{I_0}_\beta.
\end{equation}
This means that
\begin{equation}
 \na_\beta g^{I_0}_\alpha
=\df{1}{2}\epsilon_{\beta\alpha}\na^\gamma g^{I_0}_\gamma.
\end{equation}
Furthermore, 
the actions of $\na^\beta$ on both hand sides lead to 
\begin{equation}
 \na^2 g^{I_0}_\alpha=0.
\end{equation}
Similarly, for $B=\d\beta$, $A=\d\alpha$, we obtain 
\begin{equation}
 \b\na_{\d\beta} g^{I_0}_{\d\alpha}
=-\df{1}{2}\epsilon_{\d\beta\d\alpha} \b\na_{\d\gamma}
 g^{I_0 \d\gamma},
\end{equation} 
\begin{equation}
 \b\na^2g^{I_0}_{\d\alpha}=0.
\end{equation}
These consequences suggest that we may impose the constraints
\begin{equation}
 g_\alpha^{I_0}=\lambda \na_\alpha \Psi^{I_0},
\quad
g_{\d\alpha}^{I_0}=\lambda^* \b\na_{\d\alpha}\Psi^{I_0},
\label{eq:0def1}
\end{equation}
where we took $\lambda$ as a complex constant,
and $\Psi^{I_0}$ are real primary superfields.

For $B=\beta$, $A=\d\alpha$, Eq.~\eqref{eq:0B} is 
\begin{equation}
0= \na_\beta g_{\d\alpha}^{I_0}+\b\na_{\d\alpha} g_{\beta}^{I_0}
+T_{\beta\d\alpha}{}^c g^{I_0}_c.
\end{equation}
If we take $\lambda=i$,
this equation reproduces the results in Ref.~\cite{bib:BBLR}.
In this choice, $ g_{\alpha\d\alpha}^{I_0}$ are written as
 \begin{equation}
 g_{\alpha\d\alpha}^{I_0}
=\df{1}{2}[\na_\alpha,\b\na_{\d\alpha}]\Psi^{I_0}.
\label{eq:0B3}
\end{equation}
This equation is equivalently written as in  Eq.~\eqref{eq:0B1}.
Eq.\eqref{eq:0B3} means that 
$\Psi^{I_0}$ contain the field strengths of 0-form
gauge fields in the vector components.

For $B=\d\beta$, $A=\alpha\d\alpha$, Eq.~\eqref{eq:0B} is 
\begin{equation}
0= \b\na_{\d\beta} g_{\alpha\d\alpha}^{I_0}
-\na_{\alpha\d\alpha} g_{\d\beta}^{I_0}
+(q^{(0)}\cdot F_{\d\beta,\alpha\d\alpha})^{I_0}.
\end{equation}
Using Eqs.~\eqref{eq:0def1}, \eqref{eq:0B1}
and the identity 
$ \na_\alpha \b\na^2 +4i\na_{\alpha\d\gamma}\b\na^{\d\gamma}
=\b\na^2\na_\alpha -8{\cal W}_\alpha $, 
we obtain the former equation in Eq.~\eqref{eq:0B2}.
Similarly for $B=\beta$, $A=\alpha\d\alpha$, 
the latter equation in Eq.~\eqref{eq:0B2} is obtained.

Note that the degrees of freedom between bosons and fermions
in $\Psi^{I_0}$
are matched.
If the tensor hierarchy does not exist, 
Eq.~\eqref{eq:0B2}
means 
that the higher fermion components of $\Psi^{I_0}$ vanish:
\begin{equation}
 \na^2\b\na_{\d\alpha}\Psi^{I_0}=0, \quad
\b\na^2\na_\alpha \Psi^{I_0}=0.
\label{eq:0B4}
\end{equation}
So it seems that degrees of freedom in $\Psi^{I_0}$ are 
mismatched.
In a general real superfield case, the degrees of freedom 
of the components $[C,Z,H,K,B_a,\lambda,D]$ are
 $[1,4,1,1,4,4,1]$.
In this case, Bianchi identity \eqref{eq:0B1} follows that 
vector components of $\Psi^{I_0}$ are the bosonic 
field strengths:
$[\na_\alpha,\b\na_{\d\alpha}]\Psi^{I_0}| \propto \der_m f^{I_0}|$.
Thus, vector components have only one freedom.
Then, under the constraint \eqref{eq:0B4},
degrees of freedom are
$[1,4,1,1,1,0,0]$.
So the degrees of freedom between bosons and fermions in $\Psi^{I_0}$
 are matched.
The same argument holds even if the tensor hierarchy exists.

\subsection{$D$-, $A$-, $K_A$-transformation laws}
\label{ssec:sc}
We present the 
derivations of the 
$D$-, $A$-, $K_A$-transformation laws of $(Y^{I_3},
L^{I_2},W^{I_1}_\alpha,\Psi^{I_0})$.
The transformation laws of the superfields
follow from those of 
$F^{I_p}_{A_{p+1}...A_1}$.
Since $F^{I_p}_{M_{p+1}...M_1}$ are
 invariant under $X_{\cal A'}$ transformations,
the properties are reduced to those of the
vielbein:
\begin{equation}
 \delta_G(\xi^{\cal A'}X_{\cal A'})E_B{}^M
=-E_B{}^N 
\(
\delta_G(\xi^{\cal A'}X_{\cal A'})E_N{}^C
\)
 E_C{}^M.
\end{equation}
The $D$-, $A$- and $K_A$-transformation laws of the vielbein
are obtained as follows.
\begin{itemize}
 \item $D$-transformations
\begin{equation}
  \delta_G(\xi(D)D)E_b{}^M=+\xi(D)E_b{}^M,
\quad
  \delta_G(\xi(D)D)E_{\ul\beta}{}^M=+\df{1}{2}\xi(D)E_{\ul\beta}{}^M.
\end{equation}
 \item $A$-transformations
\begin{equation}
  \delta_G(\xi(A)A)E_b{}^M=0,
\quad
  \delta_G(\xi(A)A)E_\beta{}^M=-i\xi(A)E_\beta{}^M,
\quad
  \delta_G(\xi(A)A)E^{\d\beta M}=+i\xi(A)E^{\d\beta M}.
\end{equation}
\item $S$-transformations
\begin{equation}
\begin{split}
& \delta_G(\xi(K)^{\alpha}S_{\alpha})E_b{}^M
=i E_b{}^N  E_N{}^e \xi(K)^{\delta}  (\sigma_e)_{\delta \d\gamma}
 E^{\d\gamma M}
=i \xi(K)^{\delta}  (\sigma_b)_{\delta \d\gamma}
 E^{\d\gamma M}, \\
& \delta_G(\xi(K)_{\d\alpha}\b{S}^{\d\alpha})E_b{}^M
=i E_b{}^N  E_N{}^e \xi(K)_{\d\delta}  (\b\sigma_e)^{\d\delta \gamma}
 E_\gamma{}^M
=i \xi(K)_{\d\delta}  (\b\sigma_b)^{\d\delta \gamma}
 E_\gamma {}^M.
\\
&
 \delta_G(\xi(K)^{\ul\alpha}S_{\ul\alpha})E_{\ul\beta}{}^M=0.
\end{split}
\end{equation}
\item All the 
$K_a$-transformations of the vielbein are equal to zero. 
\end{itemize}
Using these equations, the $D$-, $A$- and $K_A$-transformation
 laws 
of $(Y^{I_3}, L^{I_2},W^{I_1}_\alpha,\Psi^{I_0})$ are determined.
Note that the 
$M$-transformation laws of 
$(Y^{I_3},L^{I_2},W^{I_1}_\alpha,\Psi^{I_0})$
 are obtained by their spinor indices.

\subsubsection{3-form gauge fields}
We show the $D$-, $A$-, $K_A$-transformation laws of
$Y^{I_3}$.
$Y^{I_3}$ are given in terms of 
$\Sigma^{I_3\d\delta\d\gamma}{}_{ba}$ 
as in Eq.~\eqref{eq:Y}.
The $D$-, $A$-, $K_A$-transformations of 
$\Sigma^{I_3\d\delta\d\gamma}{}_{ba}$
are determined as follows.
\begin{itemize}
 \item $D$-transformations
\begin{equation}
  \delta_G(\xi(D)D)
 \Sigma^{I_3\d\delta\d\gamma}{}_{ba}
=  \delta_G(\xi(D)D)
E^{\d\delta Q}E^{\d\gamma P} E_b{}^N E_a{}^M \Sigma^{I_3}_{QPNM}
=3 \xi(D)  \Sigma^{I_3\d\delta\d\gamma}{}_{ba}
\label{eq:3confD}
\end{equation}
\item $A$-transformations
\begin{equation}
  \delta_G(\xi(A)A)
 \Sigma^{I_3\d\delta\d\gamma}{}_{ba}
=  \delta_G(\xi(A)A)
E^{\d\delta Q}E^{\d\gamma P} E_b{}^N E_a{}^M \Sigma^{I_3}_{QPNM}
=+2i \xi(A)  \Sigma^{I_3\d\delta\d\gamma}{}_{ba}
\label{eq:3confA}
\end{equation}
\item $S_\alpha$-transformations
\begin{equation}
\begin{split}
&\delta_G(\xi(K)^{\alpha}S_{\alpha}) 
\Sigma^{I_3\d\delta\d\gamma}{}_{ba}
\\
&
=  \delta_G(\xi(K)^{\alpha}S_{\alpha}) 
E^{\d\delta Q}E^{\d\gamma P} E_b{}^N E_a{}^M \Sigma^{I_3}_{QPNM}
\\
&
=
E^{\d\delta Q}E^{\d\gamma P} 
(
i \xi(K)^{\alpha}  (\sigma_b)_{\alpha \d\epsilon}
 E^{\d\epsilon N}
) E_a{}^M 
\Sigma^{I_3}_{QPNM}
+
E^{\d\delta Q}E^{\d\gamma P} E_b{}^N 
(
i \xi(K)^{\alpha}  (\sigma_a)_{\alpha \d\epsilon}
 E^{\d\epsilon M}
) 
\Sigma^{I_3}_{QPNM}
\\
&
=
i \xi(K)^{\alpha}  (\sigma_b)_{\alpha \d\epsilon}
\Sigma^{I_3\d\delta \d\gamma \d\epsilon}{}_a
+i \xi(K)^{\alpha}  (\sigma_a)_{\alpha \d\epsilon}
\Sigma^{I_3\d\delta \d\gamma}{}_b {}^{\d\epsilon}
\\&
=0.
\end{split}
\label{eq:3confS}
\end{equation}
\item $\b{S}^{\d\alpha}$-transformations
\begin{equation}
\begin{split}
&\delta_G(\xi(K)_{\d\alpha}\b{S}^{\d\alpha})
\Sigma^{I_3\d\delta\d\gamma}{}_{ba}
\\
&
=\delta_G(\xi(K)_{\d\alpha}\b{S}^{\d\alpha})
E^{\d\delta Q}E^{\d\gamma P} E_b{}^N E_a{}^M \Sigma^{I_3}_{QPNM}
\\
&
=
E^{\d\delta Q}E^{\d\gamma P} 
(
i \xi(K)_{\d\alpha}  (\b\sigma_b)^{\d\alpha \epsilon}
 E_\epsilon {}^N
) E_a{}^M 
\Sigma^{I_3}_{QPNM}
+
E^{\d\delta Q}E^{\d\gamma P} E_b{}^N 
(
i \xi(K)_{\d\alpha}  (\b\sigma_a)^{\d\alpha \epsilon}
 E_\epsilon {}^M
) 
\Sigma^{I_3}_{QPNM}
\\
&
=
i \xi(K)_{\d\alpha}  (\b\sigma_b)^{\d\alpha \epsilon}
\Sigma^{I_3\d\delta\d\gamma}{}_{\epsilon a}
+i \xi(K)_{\d\alpha}  (\b\sigma_a)^{\d\alpha \epsilon} 
\Sigma^{I_3\d\delta\d\gamma}{}_{b \epsilon}
\\
&
=0.
\end{split}
\label{eq:3confbS}
\end{equation}
\end{itemize}
Here, we used the constraints
 $\Sigma^{I_3\d\delta \d\gamma}{}_b
      {}^{\d\epsilon}=\Sigma^{I_3\d\delta\d\gamma}{}_{b \epsilon}=0$
in the last lines of $S_\alpha$ and $\b{S}^{\d\alpha}$
transformation laws.
 These equations lead to the 
superconformal transformation laws of $Y$ in Eq.~\eqref{eq:Yconf}.
Those of $\b{Y}$ are obtained similarly.
\subsubsection{2-form gauge fields}
The $D$-, $A$-, $K_A$-transformation laws of $L^{I_2}$ are 
obtained by the same procedure as the case of 3-form gauge fields.
We summarize the results.
\begin{itemize}
 \item $D$-transformations
\begin{equation}
  \delta_G(\xi(D)D)
H^{I_2\d\gamma}{}_{\beta a}
=2 \xi(D) H^{I_2\d\gamma}{}_{\beta a}
\end{equation}
\item $A$-transformations
\begin{equation}
  \delta_G(\xi(A)A)
H^{I_2\d\gamma}{}_{\beta a}
=0.
\end{equation}
\item $S_\alpha$-transformations
\begin{equation}
\delta_G(\xi(K)^{\alpha}S_{\alpha}) 
H^{I_2\d\gamma}{}_{\beta a}
= 
i \xi(K)^{\alpha}  (\sigma_a)_{\alpha \d\epsilon}
H^{I_2\d\gamma}_{\beta}{}^{\d\epsilon}
=0.
\end{equation}
\item $\b{S}^{\d\alpha}$-transformations
\begin{equation}
\delta_G(\xi(K)_{\d\alpha}\b{S}^{\d\alpha})
H^{I_2\d\gamma}{}_{\beta a}
= 
i \xi(K)_{\d\alpha}  (\b\sigma_a)^{\d\alpha \epsilon}
H^{I_2\d\gamma}{}_{\beta\epsilon}
=0.
\end{equation}
\end{itemize}
\subsubsection{1-form gauge fields}
The $D$-, $A$-, $K_A$-transformation laws of $W^{I_2}_\alpha$ are 
the same as the case that the tensor hierarchy does not exist.
The results are as follows.
\begin{itemize}
 \item $D$-transformations
\begin{equation}
  \delta_G(\xi(D)D)
F^{I_1\d\beta}{}_{ a}
=\df{3}{2} \xi(D)F^{I_1\d\beta}{}_{ a}
\end{equation}
\item $A$-transformations
\begin{equation}
  \delta_G(\xi(A)A)
F^{I_1\d\beta}{}_{ a}
=+i\xi(A)F^{I_1\d\beta}{}_{ a}.
\end{equation}
\item $S_\alpha$-transformations
\begin{equation}
\delta_G(\xi(K)^{\alpha}S_{\alpha}) 
F^{I_1\d\beta}{}_{ a}
= 
-i \xi(K)^{\alpha}  (\sigma_a)_{\alpha \d\epsilon}
F^{I_1\d\beta \d\epsilon}
=0.
\end{equation}
\item $\b{S}^{\d\alpha}$-transformations
\begin{equation}
\delta_G(\xi(K)_{\d\alpha}\b{S}^{\d\alpha})
F^{I_1\d\beta}{}_{ a}
= 
-i \xi(K)_{\d\alpha}  (\b\sigma_a)^{\d\alpha \epsilon}
F^{I_1\d\beta}{}_\epsilon
=0.
\end{equation}
\end{itemize}
\subsubsection{0-form gauge fields}
The $D$-, $A$-, $K_A$-transformation laws of $\Psi^{I_0}$ are 
determined as follows.
\begin{itemize}
 \item $D$-transformations
\begin{equation}
  \delta_G(\xi(D)D)
g^{I_0}_\alpha
=\df{1}{2} \xi(D)g^{I_0}_\alpha
\end{equation}
lead to 
\begin{equation}
 D \Psi^{I_0}=0.
\end{equation}
This is because
\begin{equation}
[D,\na_\alpha]= \df{1}{2}\na_\alpha.
\end{equation}
\item $A$-transformations
\begin{equation}
  \delta_G(\xi(A)A)
g^{I_0}_\alpha
=-i \xi(A)g^{I_0}_\alpha
\end{equation}
lead to 
\begin{equation}
 A \Psi^{I_0}=0.
\end{equation}
This is because
\begin{equation}
[A,\na_\alpha]=-i\na_\alpha.
\end{equation}
\item $S_\alpha$-transformations
\begin{equation}
\delta_G(\xi(K)^{\beta}S_{\beta}) g^{I_0}_\alpha=0.
\end{equation}
\item $\b{S}^{\d\alpha}$-transformations
\begin{equation}
\delta_G(\xi(K)_{\d\beta}\b{S}^{\d\beta})
g^{I_0}_{\alpha}
= 0.
\end{equation}
We need to check the consistency between
the weights of $\Psi^{I_0}$ and 
the $S_{\ul\alpha}$-invariances of $\Psi^{I_0}$.
The $S_{\ul\alpha}$-invariances of $g^{I_0}_{\ul\alpha}$
require that the weights of 
$\Psi^{I_0}$ must be $(\Delta,w)=(0,0)$ \cite{bib:KU}.
The requirements are understood as follows.
The $S_{\ul\alpha}$-transformations of 
$\na_{\ul\alpha}\Psi^{I_0}$ are generally given by
\begin{equation}
S_{\alpha}\na_\beta \Psi^{I_0}
=\epsilon_{\alpha\beta}(2D-3iA) \Psi^{I_0},
\quad
\b{S}^{\d\alpha}\b\na^{\d\beta}\Psi^{I_0}
=\epsilon^{\d\alpha\d\beta}(2D+3iA)\Psi^{I_0}.
\end{equation}
These equations lead to 
the conditions for
the $S_{\ul\alpha}$-invariances of $\na_{\ul\alpha}\Psi^{I_0}$:
\begin{equation}
 D\Psi^{I_0}=0,\quad A\Psi^{I_0}=0.
\end{equation}
Actually, $\Psi^{I_0}$ satisfy the weight conditions.
Thus, the weights are consistent with 
 $S_{\ul\alpha}$-invariances of $g_{\ul\alpha}^{I_0}$.
\end{itemize}

\section{The explicit forms of bosonic field strengths}
\label{sec:bfs}

In this appendix, we summarize the 
explicit forms of the bosonic field strengths.

For 3-form gauge fields, the double bar projections
of $\Sigma^{I_3}$ lead to the following relations
\begin{equation}
\begin{split}
&
\df{1}{4!} dx^m\wed dx^n \wed dx^p \wed dx^q \Sigma^{I_3}_{qpnm}|
\\
&
=\df{1}{4!} dx^m\wed dx^n \wed dx^p \wed dx^q 
E_m{}^A E_n{}^B E_p{}^C E_q{}^D
\Sigma^{I_3}_{DCBA}|.
\end{split}
\end{equation}
We expand this relation, and obtain
\begin{equation}
\begin{split}
&
e_m{}^a e_n{}^b e_p{}^c e_q{}^d
\Sigma^{I_3}_{dcba}|
\\
&
=
\der_q C^{I_3}_{pnm}|
+(-1)^1 
\der_p C^{I_3}_{qnm}|
+
(-1)^2 
\der_n C^{I_3}_{qpm}|
+(-1)^3
 \der_m C^{I_3}_{qpn}|
-(q^{(3)}\cdot U_{qpnm}|)^{I_3}
\\
&
\quad
-
 \df{1}{2} 
(
e_m{}^a e_n{}^b e_p{}^c \psi_q{}^{\delta}
+
(-1)^1 
e_m{}^a e_n{}^b e_q{}^c \psi_p{}^{\delta}
\\
&\quad
\hph{
+
 \df{1}{2} 
(\quad
}
+
(-1)^2 
e_m{}^a e_p{}^b e_q{}^c \psi_n{}^{\delta}
+
(-1)^3 
e_n{}^a e_p{}^b e_q{}^c \psi_m{}^{\delta}
)
\(-\df{1}{16}\)(\sigma^e)_{\delta\d\zeta}\epsilon_{ecba}
\b\na^{\d\zeta} \b{Y}^{I_3}|
\\
&
\quad
-
 \df{1}{2} 
(
e_m{}^a e_n{}^b e_p{}^c \b\psi_{q \d\delta}
+
(-1)^1 e_m{}^a e_n{}^b e_q{}^c \b\psi_{p \d\delta}
\\
&\quad
\hph{
+
 \df{1}{2} 
(\quad
}
+
(-1)^2 e_m{}^a e_p{}^b e_q{}^c \b\psi_{n \d\delta}
+
(-1)^3 e_n{}^a e_p{}^b e_q{}^c \b\psi_{m \d\delta}
)
\(+\df{1}{16}\)
(\b\sigma^e)^{\d\delta\zeta} \epsilon_{ecba} \na_\zeta Y^{I_3}|
\\
&
\quad
-
\df{1}{2}  \cdot \df{1}{2}
(
e_m{}^a e_n{}^b \psi_p{}^{\gamma} \psi_q{}^{\delta}
+
(-1)^1
e_m{}^a e_p{}^b \psi_n{}^{\gamma} \psi_q{}^{\delta}
+
(-1)^2
e_m{}^a e_q{}^b \psi_n{}^{\gamma} \psi_p{}^{\delta}
\\
&
\quad
\hph{
+
 \df{1}{2}  \cdot \df{1}{2}
(\quad
}
+
(-1)^2
e_n{}^a e_p{}^b \psi_m{}^{\gamma} \psi_q{}^{\delta}
+
(-1)^{2+2}
e_n{}^a e_q{}^b \psi_p{}^{\gamma} \psi_m{}^{\delta}
+
(-1)^{2+2}
e_p{}^a e_q{}^b \psi_m{}^{\gamma} \psi_n{}^{\delta}
)
\df{1}{2}(\sigma_{ba}\epsilon)_{\delta\gamma} \b{Y}^{I_3}|
\\
&
\quad
-
\df{1}{2}  \cdot \df{1}{2}
(
e_m{}^a e_n{}^b \b\psi_{p\d\gamma} \b\psi_{q\d\delta}
+
(-1)^1
e_m{}^a e_p{}^b  \b\psi_{n\d\gamma}  \b\psi_{q\d\delta}
+
(-1)^2
e_m{}^a e_q{}^b \b\psi_{n\d\gamma}  \b\psi_{p\d\delta}
\\
&
\quad
\hph{
+
 \df{1}{2}  \cdot \df{1}{2}
(\quad
}
+
(-1)^2
e_n{}^a e_p{}^b \b\psi_{m\d\gamma}  \b\psi_{q\d\delta}
+
(-1)^{2+2}
e_n{}^a e_q{}^b  \b\psi_{p\d\gamma}  \b\psi_{m\d\delta}
+
(-1)^{2+2}
e_p{}^a e_q{}^b  \b\psi_{m\d\gamma}  \b\psi_{n\d\delta}
)
(\b\sigma_{ba}\epsilon)^{\d\delta\d\gamma} Y^{I_3}|
\end{split}
\end{equation}
For the 2-form gauge fields, 
the double bar projections are
\begin{equation}
H^{I_2}||
= \df{1}{3!}dx^m \wed dx^n \wed dx^p H^{I_2}_{pnm}|
=\df{1}{3!} dx^m \wed dx^n \wed dx^p E_m{}^A E_n{}^B E_p{}^C H^{I_2}_{CBA}|.
\end{equation}
We obtain the component expressions of 
bosonic field strengths
\begin{equation}
\begin{split}
& e_m{}^a e_n{}^b e_p{}^c H^{I_2}_{cba}|
\\
&
=
\der_p B^{I_2}_{nm}|+\der_n B^{I_2}_{mp}|+\der_m B^{I_2}_{pn}|
-(q^{(2)}\cdot C_{pnm}|)^{I_2}
\\
&
\quad
-
 \df{1}{2}
(
e_m{}^a e_n{}^b \psi_p{}^{{\gamma}} 
+
(-1)^1 e_m{}^a \psi_n{}^{\gamma} e_p{}^b 
+
(-1)^3\psi_m{}^{\gamma} e_n{}^b e_p{}^a
)
(+2)(\sigma_{ba})_\gamma{}^\delta \na_\delta L^{I_2}|
\\
&
\quad
-
 \df{1}{2}
(
e_m{}^a e_n{}^b \b\psi_{p\d{\gamma}} 
+
(-1)^1 e_m{}^a \b\psi_{n\d\gamma} e_p{}^b 
+
(-1)^3 \b\psi_{m\d\gamma} e_n{}^b e_p{}^a
)
(+2)(\b\sigma_{ba})^{\d\gamma}{}_{\d\delta}\b\na^{\d\delta}L^{I_2}|
\\
&
\quad
-
\df{1}{2} \cdot \df{1}{2}
(
 e_m{}^a \psi_n{}^{\beta} \b\psi_p{}^{\d\gamma}
+
(-1)^1
\psi_m{}^{\beta} e_n{}^a \b\psi_p{}^{\d\gamma} 
+
(-1)^3
\psi_n{}^{\beta} \b\psi_m{}^{\d\gamma} e_p{}^a 
)
(-1)
(+2i)
(\sigma_a)_{\beta\d\gamma}L^{I_2}|
\\
&
\quad
-
\df{1}{2} \cdot \df{1}{2}
(
 e_m{}^a \b\psi_{n\d\beta} \psi_{p\gamma}
+
(-1)^1
\b\psi_{m\d\beta} e_n{}^a \psi_{p\gamma} 
+
(-1)^3
\b\psi_{n\d\beta} \psi_{m\gamma} e_p{}^a 
)
(-1)
(+2i) (\b\sigma_a)^{\d\beta\gamma} L^{I_2}|.
\end{split}
\end{equation}
For 1-form gauge fields, the double bar projections are
\begin{equation}
  F^{I_1}||
=\df{1}{2!} dx^m\wed dx^n F^{I_1}_{nm}|
=\df{1}{2!} dx^m\wed dx^n E_m{}^AE_n{}^BF_{BA}^{I_1}|.
\end{equation}
We obtain the expressions of the bosonic field strengths
\begin{equation}
\begin{split}
e_m{}^a e_n{}^b F_{ba}^{I_1}|
&=
\der_n A^{I_1}_m |-\der_m A^{I_1}_n|-(q^{(1)}\cdot B_{nm}|)^{I_1}
\\
&
\quad
-
\df{1}{2}
(
e_m{}^c\psi_n{}^{\beta}
-e_n{}^c\psi_m{}^{\beta}
)(-1)(\sigma_{c})_{\beta\d\gamma}\b{W}^{I_1\d\gamma}|  
\\
&
\quad
-
\df{1}{2}
(
e_m{}^c\b\psi_{n\d\beta}
-e_n{}^c\b\psi_{m\d\beta}
)
(+1)(\b\sigma_c)^{\d\beta\gamma}W^{I_1}_\gamma|.
\end{split}
\end{equation}
The above expressions are basic building blocks in the 
constructions of component field actions. 

\end{document}